# Experimental and theoretical study of $SbPO_4$ under compression

**André Luis de Jesus Pereira,[1,2*] David Santamaría-Pérez,[3] Rosário Vilaplana,[4] Daniel Errandonea,[3] Catalin Popescu,[5] Estelina Lora da Silva,[1] Juan Angel Sans,[1] Juan Rodríguez-Carvajal,[6] Alfonso Muñoz,[7] Plácida Rodríguez-Hernández,[7] Andres Mujica,[7] Silvana Elena Radescu,[7] Armando Beltrán,[8] Alberto Otero de la Roza,[9] Marcelo Nalin,[10] Miguel Mollar,[1] and Francisco Javier Manjón[1*]**

[1] *Instituto de Diseño para la Fabricación y Producción Automatizada, MALTA Consolider Team, Universitat Politècnica de València, València, Spain*

[2] *Grupo de Pesquisa de Materiais Fotonicos e Energia Renovável - MaFER, Universidade Federal da Grande Dourados, Dourados, MS, Brazil*

[3] *Departament de Física Aplicada – ICMUV, MALTA Consolider Team, Universitat de València, Burjassot, Spain*

[4] *Centro de Tecnologías Físicas, MALTA Consolider Team, Universitat Politecnica de València, València 46022, Spain*

[5] *CELLS-ALBA Synchrotron Light Facility, 08290 Cerdanyola, Barcelona, Spain*

[6] *Institut Laue-Langevin, 71 Avenue des Martyrs, CS 20156, 38042, Grenoble, Cedex 9, France*

[7] *Departamento de Física, Instituto de Materiales y Nanotecnología, MALTA Consolider Team, Universidad de La Laguna, Tenerife, Spain*

[8] *Departament de Química Física i Analítica, MALTA Consolider Team, Universitat Jaume I, Castelló, Spain*

[9] *Departamento de Química Física y Analítica, MALTA Consolider Team, Universidad de Oviedo, 33006 Oviedo, Spain*

[10] *Instituto de Quimica, Departamento de Química Geral e Inorgânica, UNESP - Campus de Araraquara, SP, Brazil*

## Abstract

$SbPO_4$ is a complex monoclinic layered material characterized by a strong activity of the non-bonding lone electron pair (LEP) of Sb. The strong cation LEP leads to the formation of layers piled up along the *a*-axis and linked by weak Sb-O electrostatic interactions. In fact, Sb is 4-fold coordination with O similar to what occurs with the P-O coordination, despite the large difference of ionic radii and electronegativity between both elements. Here we report a joint experimental and theoretical study of the structural and vibrational properties of $SbPO_4$ at high pressure. We show that $SbPO_4$ is not only one of the most compressible phosphates but also one of the most compressible compounds of the $ABO_4$ family. Moreover, it has a considerable anisotropic compression behavior with the largest compression occurring along a direction close to *a*-axis and governed by the compression of the LEP and the weak inter-layer Sb-O bonds. The strong compression along the *a*-axis leads to a subtle modification of the monoclinic crystal structure above 3 GPa leading from a 2D to a 3D material. Moreover, the onset of a reversible pressure-induced phase transition is observed above 9 GPa, which is completed above 20 GPa. We propose that the high-pressure phase is a triclinic distortion of the original monoclinic phase. The understanding of the compression mechanism of $SbPO_4$ can aid in understanding the importance of the ion intercalation and catalytic properties of this layered compound.

* Corresponding authors, email: andrepereira@ufgd.edu.br, fjmanjon@fis.upv.es

## 1. Introduction

Inorganic functional materials composed by antimony, such as antimony orthophosphate ($SbPO_4$), are receiving considerable attention from the scientific community, due to their potential applications in different areas. The excellent optical properties of antimony-based glasses, such as the high-linear refractive index[1,2] and the large transmittance window from ultraviolet (UV) to infrared (IR) regions[3] enables its application as optical fibers, allowing its use in photonic applications[4]. The remarkable optical properties of $SbPO_4$ also drew a lot of attention as a photocatalyst under UV light irradiation[5,6]. Moreover, since $SbPO_4$ belongs to a class of phosphates with a very stable layered structure, where ions can be intercalated between its layers, many research groups have studied its ion-exchange characteristics[7] and respective potential as anode in lithium-ion batteries[8,9].

$SbPO_4$ is an $A^{3+}B^{5+}O_4$ compound with layered structure that crystallizes in the monoclinic $P2_1/m$ (No. 11) space group, which is isostructural to $SbAsO_4$[10] and belongs to the same space group as the polymorph $BiPO_4$-III[11,12]. The low-pressure (LP) structure of $SbPO_4$ is composed by a combination of regular $PO_4$ tetrahedra and $SbO_4E$ polyhedra disposed in a trigonal bipyramidal fashion, where E refers to the non-bonding lone electron pair (LEP) of Sb (see **Figure 1(a)**). For both $BiPO_4$-III and $SbPO_4$, P is 4-fold coordinated at room pressure; however, while $SbPO_4$ is a layered compound, $BiPO_4$-III is not, therefore these are not isostructural compounds. At room pressure, the Bi ion belonging to the $BiPO_4$-III compound features a 6-fold coordination, whereas Sb has only a 4-fold coordination for $SbPO_4$. The difference between both compounds is the result of the stronger LEP of Sb(III) when compared to the LEP of Bi(III), i.e. the strong Sb LEP prevents the formation of Sb-O bonds in either direction, thus leading to the formation of layers in $SbPO_4$ unlike what occurs for $BiPO_4$-III. The layers of $SbPO_4$ are piled up along the *a*-axis and are linked by weak Sb-O electrostatic interactions.

The vibrational properties of $SbPO_4$ have been studied at room pressure by Raman and IR spectroscopy[13–16], but a limited amount of information has been provided. For instance, the classification and symmetry assignment of all vibrational modes at the Brillouin zone (BZ) center (Γ), the phonon dispersion curves (PDCs) and the phonon density of states (PDOS) have not been reported even at room pressure.

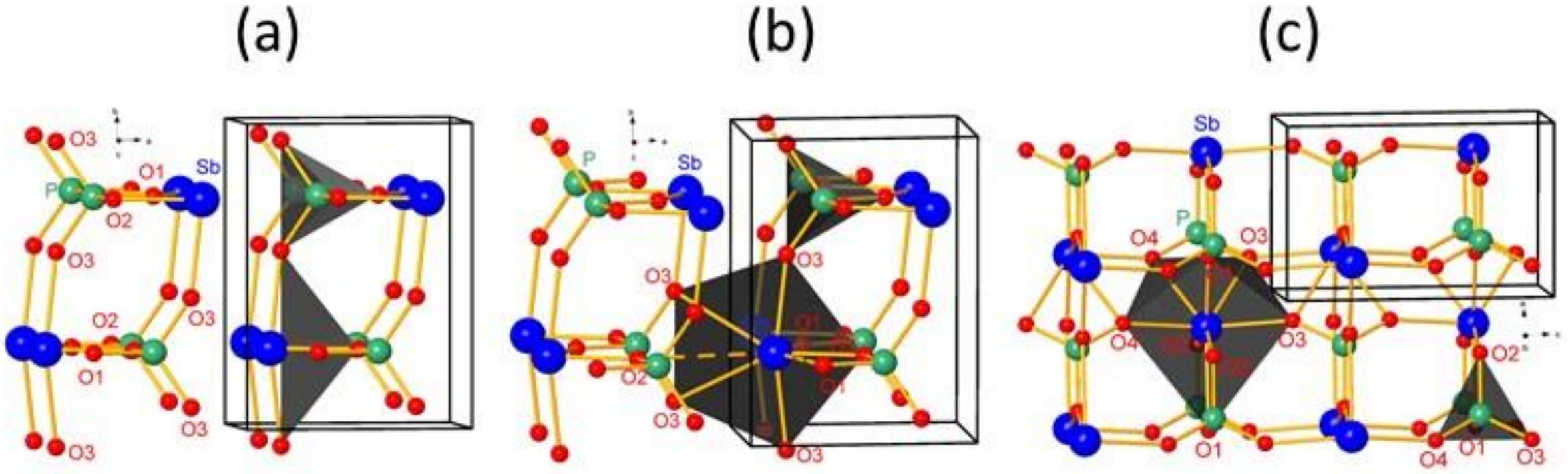

**Figure 1.** (a) 2D crystalline structure of monoclinic $SbPO_4$ at room pressure. The layered structure is composed of P atoms which a 4-fold coordinated to O in the $sp^3$ tetrahedral arrangement and Sb atoms which are 4-fold coordinated to O in a trigonal bipyramid fashion. The distortion of the Sb polyhedron is caused by the strong LEP which points to the interlayer space (not shown). (b) 3D crystalline structure of monoclinic $SbPO_4$ at 3 GPa, showing the 4 + 2-fold coordination for Sb if one includes not only four Sb-O distances below 2.2 Å, but also two Sb-O distances just below 2.6 Å (see Figure 6). Dashed lines show other two Sb-O distances above 2.8 Å that do not enter the Sb coordination. The position of the Sb LEP can be considered intermediate between the two dashed lines. (c) 3D crystalline structure of triclinic $SbPO_4$ at 18 GPa showing a 4+2+1-fold coordination for Sb in a distorted dodecahedral fashion (Sb-O distances are between 2.0 and 2.7 Å) and 4-fold coordinated P atoms in tetrahedral $sp^3$ arrangement. Blue big balls represent Sb atoms, green medium balls represent P atoms and red small balls represent O atoms.

High-pressure (HP) studies of several *A*$PO_4$ orthophosphates have been reported in the literature, i.e. where zircon- and monazite-type phosphates have been broadly studied[17–27]. The pressure-induced structural sequence has been understood, with several new HP phases discovered and their crystal structures solved. In addition, the influence of pressure in the vibrational properties and unit-cell parameters has been well established. The number of HP studies carried out has also helped to unveil the existing relationship between the response under compression of the microscopic and macroscopic properties of these materials. In particular, the compressibility has been explained in terms of polyhedral compressibilities[23]. Moreover, the studies of phosphates under extreme conditions have been recently extended to compounds with different crystal structures when compared to zircon or monazite. In particular, phosphates with the olivine structure as well as complex phosphates, like $K_2Ce(PO_4)_2$ and isomorphic compounds, have been characterized under HP[28,29]. Finally, metastable polymorphs of $BiPO_4$[18], spin-Peierls distorted $TiPO_4$[30] and $CrVO_4$–type phosphates[31] have also been recently studied at HP. Phase transitions (PTs) driven by compression have been reported for all these compounds, with a common feature found that the PTs are always first-order, involving a collapse of the volume and the breaking and formation of chemical bonds. Moreover, HP has been found to be a successful route to penta-coordinated phosphorus, which is achieved at a pressure of 46 GPa in $TiPO_4$[30]. In contrast with all the phosphates mentioned above, the HP behavior of $SbPO_4$ has not yet been explored. Being this structure a layered compound and Sb possessing a strong LEP, $SbPO_4$ is an ideal candidate for an unusual HP behavior with high compressibility and with subtle PTs at much lower pressure than that found for other $APO_4$ orthophosphates.

In this work, we report a joint experimental and theoretical study of the structural and vibrational properties of $SbPO_4$ at HP by means of x-ray diffraction (XRD) and Raman scattering (RS) measurements combined with *ab-initio* calculations. We will show that $SbPO_4$ is one of the most compressible phosphates and $ABO_4$ compounds. Moreover, it exhibits a considerable anisotropic behavior due to a high non-linear compression, mainly along the *a*-axis, as shown by respective compressibility tensor. Additionally, we will show that our measurements and calculations are compatible with the existence of an isostructural phase transition (IPT) around 3 GPa and a reversible PT above 9 GPa, which is completed around 20 GPa. After the study of several candidates for the HP phase of $SbPO_4$ based on an updated Bastide's diagram[20,32] for $ABO_4$ compounds containing cations with LEPs, like $As^{3+}$, $Sb^{3+}$, $Bi^{3+}$, $Sn^{2+}$ and $Pb^{2+}$, we propose a triclinic distortion of the original monoclinic phase as the HP phase above 9 GPa. The experimental and theoretical vibrational modes of both LP and HP phases at different pressures will be shown and a tentative assignment of the symmetry of each observed Raman-active mode will be provided. This work helps to better understand how layered $SbPO_4$ behaves under compression and provides clues to design better photocatalysts and better intercalated compounds with enhanced ion-exchange characteristics based on this phosphate. This work helps us understand the behavior of the layered $SbPO_4$ under compression, thus providing insights to direct and improve the design of similar photocatalysts and intercalated compounds with enhanced ion-exchange characteristics.

## 2. Experimental Method

Synthetic $SbPO_4$ powders used in the present experiments were synthesized by M. Nalin and coworkers[2–4]. Energy dispersive X-ray spectroscopy (EDS) analyses performed with an Oxford Instruments detector coupled to a JEOL JSM6300 scanning electron microscope showed a good stoichiometry and no appreciable impurities.

Structural characterization of powders at room pressure was carried out by XRD measurements performed with a Rigaku Ultima IV diffractometer using Cu $K_\alpha$ (1.5406 and 1.5443 Å for $K_{\alpha1}$ and $K_{\alpha2}$, respectively) as the incident radiation source. Traces of other phases or of $Sb_2O_3$ were not detected. Vibrational characterization of powders at room pressure was carried out by RS measurements performed with a Horiba Jobin Yvon LabRAM HR UV microspectrometer, equipped with a thermoelectrically cooled multichannel charge-coupled device detector and a 1200 grooves/mm grating that allows a spectral resolution better than 3 $cm^{-1}$. The signal was collected in backscattering geometry exciting with a 532 nm laser with a power of less than 10 mW. Phonons were analyzed by fitting Raman peaks with a Voigt profile fixing the Gaussian linewidth (2.4 $cm^{-1}$) to the experimental setup resolution. RS experiments allowed us also to confirm that the samples contained only a pure phase.

Powder angle-dispersive HP-XRD measurements were performed at room temperature in three different experiments. Initially, we performed two experiments (called Run 1 and Run 2) using an Xcalibur diffractometer with the lines $K_{\alpha1}$ and $K_{\alpha2}$ of a Mo source with $\lambda = 0.7093$ and 0.7136 Å, respectively. The sample was loaded with a 16:3:1 methanol-ethanol-water mixture in a Merrill-Bassett-type diamond anvil cell (DAC) with diamond culets of 400 μm in diameter[33]. A third powder angle-dispersive HP-XRD experiment (Run 3) was performed up to 15.2 GPa in the BL04-MSPD beamline at ALBA synchrotron facility[34]. This beamline is equipped with Kirkpatrick-Baez mirrors to focus the monochromatic beam and a Rayonix CCD detector with a 165 mm diameter-active area and was operated with a wavelength of 0.4246 Å. In the two first experiments, pressure was determined by the luminescence of small ruby chips evenly distributed in the pressure chamber[35], while in the third experiment pressure was determined with the equation of state (EoS) of copper[36]. Integration of 2D diffraction images was performed with Dioptas software[37] while structural analysis was performed by Rietveld and Le Bail refinements using FullProf[38] and PowderCell[39] program packages. In all the experiments the DAC loading was performed taking care of avoiding sample bridging with the gasket[40].

Finally, unpolarized HP-RS measurements up to 24.5 GPa were performed with the Horiba Jobin Yvon LabRAM HR UV microspectrometer previously mentioned. The sample was loaded with a 16:3:1 methanol-ethanol-water mixture in a membrane-type DAC and pressure was determined by the ruby luminescence method[35]. In the pressure range covered by Raman and XRD experiments pressure was determined with an accuracy of 0.1 GPa.

## 3. Theoretical details

*Ab-initio* calculations were performed within the framework of density functional theory (DFT)[41] to study the structural, vibrational, and elastic properties of $SbPO_4$ under pressure. Simulations were carried out with the Vienna *ab-initio* simulation package (VASP)[42] using the projector augmented wave (PAW) pseudopotentials[43]. The PAW scheme replaces core electrons by smoothed pseudovalence wave functions considering the full nodal character of the all-electron charge density in the core region. The set of plane waves employed was extended up to a kinetic energy cutoff of 520 eV because of the presence of oxygen in $SbPO_4$. The generalized gradient approximation (GGA) was used for the description of the exchange-correlation energy within the PBEsol prescription[44]. The BZ of the monoclinic and the others analyzed structures of $SbPO_4$ were sampled with dense Monkhorst-Pack grids of special **k**-points[45]. A high convergence of 1-2 meV per formula unit in the total energy is achieved with the cutoff energy and the **k**-point sampling employed. This ensures an accurate calculation of the forces on atoms. At a set of selected volumes, the structure was fully relaxed to the optimized configuration through the calculation of the forces on the atoms and the stress tensor until the forces on the atoms were smaller than 0.005 eV/Å and the deviations of the stress tensor from a diagonal hydrostatic form were lower than 0.1 GPa.

Lattice-dynamics calculations were performed to study the phonons at the **Γ**-point of the BZ using the direct force constant approach (or supercell method). The diagonalization of the dynamical matrix provides the frequency and symmetry of the phonon modes. In order to obtain the PDCs along high-symmetry directions of the BZ and the PDOS, similar calculations were performed using appropriate supercells (2x2x2), which allow the PDCs at **k**-points to be obtained commensurate with the supercell size[46]. Finally, in order to study the HP mechanical stability of $SbPO_4$, the elastic stiffness constants were determined employing the stress theorem[47]. The optimized structures were strained, at different pressures, considering their symmetry[48].

In order to analyze the Sb-O interatomic interactions of $SbPO_4$ at different pressures, we computed the electron density and its Laplacian at the Sb-O bond critical points using the VASP code and the CRITIC2 program[49]. The CRITIC2 code implements the Quantum Theory of Atoms in Molecules (QTAIM)[50]. Within this theory, the 1-saddle critical points of the electron density (bond critical points, BCPs), and their corresponding atomic interaction lines (bond paths), determine which atoms are bonded to which. In addition, the value of ρ at the BCP correlates with the strength of the bond between two nuclei, provided the comparison is restricted to pairs of atoms of the same species. The Laplacian of the charge density at the BCP, $\nabla^2\rho(\mathbf{r}) = 0$, can be used to determine the covalent (if $\nabla^2\rho(\mathbf{r}) < 0$) character of the bond. We note that the charge density computed from the present PAW-DFT calculations using VASP solely contains the valence states; consequently, the calculated charge density values are only relevant for distances larger than the PAW radius of each atom (far enough from the core). However, it is this region where the BCPs appear and therefore the analysis of the density can still be used to characterize the Sb-O bond. We also performed an analysis of the electron localization function (ELF) along the Sb-O bonds. For the ELF analysis, we used the Elk software[51] version 6.3.2 with structural parameters obtained from the VASP optimization. The Elk software provides all-electron full-potential linearized augmented plane-wave (FPLAPW) calculations. We used a 4x4x4 uniform grid for the reciprocal space sampling, a Rmin*Kmax equal to 7.0, and a Gmax for the interstitial expansion of the density and potential equal to 22.0 a.u. To have smoother ELF profiles, we increased the number of radial points inside the muffin tins to 1000, except in the 20.8 GPa case, where this causes SCF convergence difficulties.

## 4. Results

### 4.1. Structural and vibrational properties at ambient conditions

The XRD diffractogram of $SbPO_4$ at room pressure is shown in **Figure 2(a)**. Rietveld refinement of the XRD pattern was performed using, as initial model, the monoclinic $P2_1/m$ (space group No. 11) structure of $SbPO_4$ reported in literature[9]. The refined parameters were the overall scale factor, the zero shift, the cell parameters, the pseudo-Voigt profile function with terms to account for the reflection anisotropic broadening (including anisotropic micro-strains), the fractional atomic coordinates, and the background. The Rietveld refinement yielded the following lattice parameters at 1 atm: $a$ = 5.10303(4) Å, $b$ = 6.77210(3) Å, $c$ = 4.74424(3) Å, $\beta$ = 94.6089(4)º, a unit-cell volume $V_0$ = 163.422(2) $Å^3$, and the atomic coordinates collected in **Table 1**. These values agree with values reported in the literature: $a$ = 5.0868, $b$ = 6.7547 Å, $c$ = 4.7247 Å, β = 94.66º and $V_0$ = 161.8 $Å^3$ [11]. Our experimental values agree with those from our own *ab-initio* calculations (see **Table S1** in Supplementary Information (SI)). We have found that the theoretical $V_0$ underestimates the experimental $V_0$ by only a 0.4%; a value that is within the uncertainty in GGA-PBESol calculations.

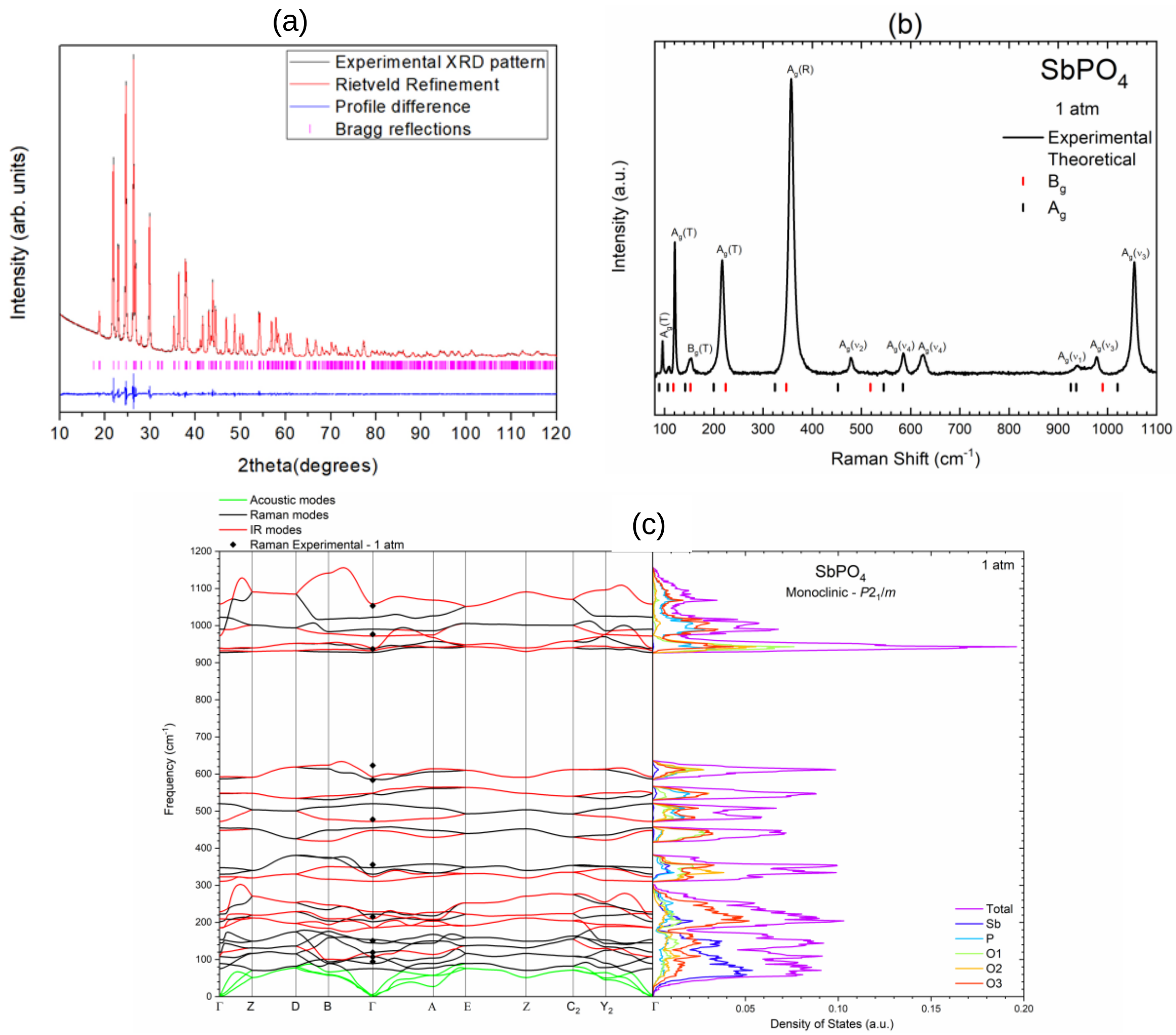

**Figure 2.** Characterization of monoclinic $SbPO_4$ at room pressure: Experimental powder XRD pattern (a) and RS spectrum (b) and theoretical PDC and PDOS (c). Bottom marks in (a) show the angle of diffraction peaks and residuals of Rietveld refinement of the XRD data. In (b) we have added a tentative mode assignment based in the theoretical results and the pressure evolution of these vibrational modes. Bottom marks in (b) show the frequencies of the Raman-active modes theoretically predicted at 0 GPa. Notations T, R, $\nu_1$, $\nu_2$, $\nu_3$, and $\nu_4$ refer to the main character of translation, rotation, or internal modes of the $PO_4$ units, respectively.

At room pressure, the monoclinic structure of $SbPO_4$ has one Sb, one P and three O atoms at independent Wyckof sites: all atoms are at 2*e* (x,1/4,z) sites except one O that is at a 4*f* (x,y,z) sites. Therefore, the monoclinic structure has eleven free atomic parameters. For this structure, P atoms are 4-fold coordinated by O atoms and form regular tetrahedra with the P atom in the center and with P-O bond lengths ranging from 1.509(4) Å to 1.536(6) Å. On the other hand, Sb atoms are 4-fold unilaterally coordinated by O atoms and form a $SbO_4E$ polyhedra in a trigonal bipyramidal fashion with the Sb-O bond lengths ranging from 2.000(6) to 2.193(3) Å (see **Figure 1(a)**). $PO_4$ tetrahedra and $SbO_4E$ polyhedra are connected by their edges, so each $PO_4$ unit is linked to four $SbO_4E$ units and each $SbO_4E$ unit is linked to four $PO_4$ units.

The $PO_4$ and $SbO_4E$ units are repeated along the [010] and [001] directions, forming layers along the [100] direction. As it was observed on other Sb-based and Bi-based sesquioxides and sesquichalcogenides[52–56], the presence of a strong non-bonding cation LEP in $SbPO_4$ causes a distortion in the structure that usually leads to a layered structure.

**Table 1.** Atomic coordinates of monoclinic $P2_1/m$ (space group No. 11) structure of $SbPO_4$ at ambient conditions obtained by Rietveld refinement. Lattice parameters are: $a$ = 5.10303(4) Å, $b$ = 6.77210(3) Å, $c$ = 4.74424(3) Å and β = 94.6089(4)º, with a unitcell volume $V_0$ = 163.422(2) $Å^3$.

| Atom | Wyckoff position | *x* | *y* | *z* |
|---|---|---|---|---|
| Sb | 2e | 0.18091(13) | 0.25 | 0.20633(14) |
| P | 2e | 0.6120(4) | 0.25 | 0.7215(5) |
| O1 | 2e | 0.3382(10) | 0.25 | 0.8346(10) |
| O2 | 2e | 0.5520(8) | 0.25 | 0.3997(9) |
| O3 | 4f | 0.7714(7) | 0.0712(4) | 0.8179(7) |

As regards the lattice dynamics of $SbPO_4$, **Figure 2(b)** shows the experimental RS spectrum of $SbPO_4$ observed at room pressure. The RS spectrum accounts for 13 peaks at room pressure and is dominated by a strong mode close to 356 $cm^{-1}$. In fact, our RS spectrum is similar to the only one that has been published up to our knowledge[15] and is similar in appearance to that of $BiPO_4$-III[12]. Group theoretical considerations of the $P2_1/m$ structure yield 36 normal modes of vibration at **Γ**, whose mechanical decomposition is[57]:

$$\Gamma = 11\ A_g(R) + 7\ B_g(R) + 6\ A_u(IR) + 9\ B_u(IR) + A_u + 2\ B_u$$

where $A_g$ and $B_g$ modes are Raman-active (R) and $A_u$ and $B_u$ are infrared-active (IR), except for one $A_u$ and two $B_u$ modes that are the three acoustic modes. Therefore, the 33 optical modes are divided in eighteen Raman-active modes ($\Gamma_{Raman}$= 11 $A_g$ + 7 $B_g$) and fifteen IR-active modes ($\Gamma_{IR}$= 6 $A_u$ + 9 $B_u$). is the same proposed in **Ref. 12** for $BiPO_4$-III with the same space group but different from that proposed in **Ref. 15**. Marks at the bottom of **Figure 2(b)** show the theoretical frequencies of $SbPO_4$ at 0 GPa, for comparison with experimental data. A tentative assignment of the symmetry of the experimental Raman-active modes on the light of our theoretical calculations is provided in **Table 2**.

**Table 2.** Experimental and theoretical Raman mode frequencies at zero pressure and pressure coefficients of $SbPO_4$ as obtained by fitting the equation $\omega(P) = \omega_0 + a \cdot P$ up to 3 GPa. Notations T, R, $\nu_1$, $\nu_2$, $\nu_3$, and $\nu_4$ refer to the main character of translation, rotation, or internal modes of the $PO_4$ units.

| | Experimental | | Theoretical | |
|---|---|---|---|---|
| Symmetry | $\omega_0$ ($cm^{-1}$) | $a$ ($cm^{-1}$/GPa) | $\omega_0$ ($cm^{-1}$) | $a$ ($cm^{-1}$/GPa) |
| $B_g$ (R) | | | 75 | 4.5 |
| $A_g$ (T) | 96 | 0.9 | 89 | 0.5 |
| $A_g$ (T) | 121 | -0.4 | 106 | -0.6 |
| $B_g$ (R) | | | 118 | 5.1 |
| $A_g$ (T) | | | 142 | 2.6 |
| $B_g$ (T) | 152 | 1.7 | 152 | 2.3 |
| $A_g$ (T) | 217 | 0.3 | 200 | 0.6 |
| $B_g$ (R) | | | 224 | 6.1 |
| $A_g$ (R) | 357 | 1.4 | 324 | -1.2 |
| $B_g$ ($\nu_2$) | | | 347 | -0.8 |
| $A_g$ ($\nu_2$) | 478 | 3.1 | 452 | 4.4 |
| $B_g$ ($\nu_4$) | | | 518 | 2.4 |
| $A_g$ ($\nu_4$) | 584 | -0.5 | 545 | -2.3 |
| $A_g$ ($\nu_4$) | 623 | 0.9 | 584 | 1.7 |
| $A_g$ ($\nu_1$) | 937 | 6.0 | 926 | 7.8 |
| $A_g$ ($\nu_3$) | 977 | 3.6 | 937 | 5.4 |
| $B_g$ ($\nu_3$) | | | 990 | 0.5 |
| $A_g$ ($\nu_3$) | 1053 | 4.0 | 1020 | 6.2 |

**Figure 2(c)** shows the theoretical PDCs and PDOS at 0 GPa. Acoustic, Raman- and IR-active branches close to **Γ** have been distinguished and a large phonon gap is observed between 650 and 920 $cm^{-1}$. In order to understand the vibrational modes of $SbPO_4$ at room pressure, we analyzed the **Γ** eigenvectors computed from the simulations performed at 0 GPa and the J-ICE[58] visualization software (see SI for further information). Out of the 33 optical modes, we can comment first on the shear (or transverse) and compressional (or longitudinal) rigid layer modes, which are low-frequency modes typical of layered compounds. For most layered compounds with tetragonal or hexagonal symmetry, the shear rigid layer mode is an $E_g$ mode (double degenerate)[59,60]. However, $SbPO_4$ has a monoclinic symmetry where double degenerate modes are not allowed. For this reason, two shear rigid layer modes ($B_g$ mode at 75 $cm^{-1}$ and $A_g$ mode at 89 $cm^{-1}$) are observed for $SbPO_4$ (see **Figures S1 and S2** in SI). On the other hand, the $A_g$ mode at 106 $cm^{-1}$ found for $SbPO_4$ is attributed to the compressional rigid layer mode, despite of not being possible to observe a complete movement of one layer against the other (see **Figure S3** in SI). In the shear rigid layer modes, the atomic vibrations are mainly along the different axis containing the layers (*b*- and *c*-axes for $SbPO_4$) whereas for the compressional rigid layer mode the atomic vibrations refer to the movement of one layer with the adjacent neighbor layer (Sb atoms against the opposite layer mainly along the *a*-axis). These rigid layer modes are low-frequency modes and are mostly related to movement of heavy atoms (Sb in $SbPO_4$) as observed in the PDOS below 200 $cm^{-1}$ (see **Figure 2(c)**).

It is also noteworthy of mentioning that for $SbPO_4$ the internal modes associated with the $PO_4$ tetrahedron, are bending and stretching P-O modes located on the medium and high-frequency regions, respectively (see PDOS in **Figure 2(c)**). In fact, as in many phosphates, the vibrational modes of $SbPO_4$ can be understood as internal and external modes of the $PO_4$ units. It is known that the internal modes of the free tetrahedral $PO_4^{3-}$ molecule with $T_d$ symmetry are: the symmetric stretching $A_1$ mode (aka $\nu_1$), the triply degenerated ($F_2$) asymmetric stretching (aka $\nu_3$), the doubly degenerated ($E$) bending mode (aka $\nu_2$); and the triply degenerated ($F_2$) bending mode (aka $\nu_4$). These vibrations are located at 938, 1017, 420 and 567 $cm^{-1}$, respectively[61]. In $SbPO_4$, the highest frequency modes (above 900 $cm^{-1}$) are mainly asymmetric stretching modes, except the symmetrical P-O stretching mode ($A_g$ mode of 936 $cm^{-1}$) in which the four O atoms vibrate in phase against the P atom (see **Figure S4** in SI). The medium-frequency modes between 400 and 650 $cm^{-1}$ are mostly related to P-O bending: i) above 540 $cm^{-1}$ these correspond to P-O bending modes combined with Sb-O stretching modes and ii) below 540 $cm^{-1}$ these correspond to P-O and Sb-O bending modes of both $PO_4$ and $SbO_4$ units. Therefore, we understand that the phonon gap found on $SbPO_4$ is clearly due to the separation of the internal stretching and bending modes evidenced inside the $PO_4$ units.

Finally, the vibrational modes of the low-frequency region below 400 $cm^{-1}$ can be related to translations (T) and rotations (R) of the $PO_4$ units; i.e. the external modes of the $PO_4$ units (see **Table 2** and **Table S3** in SI). In particular, the $A_u$ mode of 220 $cm^{-1}$ corresponds to the rotation of the $PO_4$ units (see **Figure S5** in SI) and other modes at frequencies between 200 and 330 $cm^{-1}$ also show partial rotation of the $PO_4$ units. It is noteworthy of mentioning that the four internal modes of the $PO_4$ units have basically the same frequency of most $APO_4$ compounds due to the strong covalent bond between P and O atoms as compared to the weaker ionic-covalent *A*-O bonds. This leads us to consider that the $PO_4$ units of most of the $APO_4$ compounds are isolated units stuffed with *A* cations that lead to a minor perturbation of P-O bonds. The similar frequencies of the internal phonons of the $PO_4$ units of the different phosphates justifies the incompressibility of the $PO_4$ tetrahedron in comparison with other polyhedral units related to the *A* cation[19]. We would like to point out that, despite the description given above, is possible to perform more elaborated analyzes of the origin of the $SbPO_4$ vibrational modes using, for example, the concept of bond stiffness[62–64].

### 4.2. Structural properties under compression

All XRD peaks shift to larger angles on increasing pressure up to 15.2 GPa as observed in **Figure S6**. This result is consistent with the decrease of interplanar distances at increasing pressure. In addition to that, from room pressure up to 8.4 GPa, the only noticeable change on the XRD pattern is the gradual increase of the intensity of the peak at the lowest angle. This phenomenon is the consequence of changes in the coordinate of Sb, which slowly moves from the room pressure position to that of Bi in $BiPO_4$-III, favoring the approximation of Sb to two second-neighboring oxygen atoms; a fact supported by our ab initio simulation. Above 8.4 GPa, we observe the progressive appearance of four additional diffraction peaks (see **Figure S6**). The new peaks increase in intensity continuously up to the maximum pressure of our XRD study and these are not related to the monoclinic $SbPO_4$ structure. On pressure release, the obtained diffraction pattern is identical to that of the initial sample, thus showing the reversibility of the pressure-induced PT (see top of **Figure S6**).

**Figure 3** shows the experimental and theoretical pressure dependence of the unit-cell volume of monoclinic $SbPO_4$ up to 14.8 GPa. A 3$^{rd}$-order Birch-Murnaghan Equation of State (BM-EoS)[65] was fitted to our *P*-*V* data to obtain the zero-pressure volume, $V_0$, bulk modulus, $B_0$, and its pressure derivative, $B_0'$. If the volume vs pressure data are fitted in the whole range, a $B_0'$ larger than 10 is obtained, thus suggesting an anomalous compressibility behavior. As will be commented below, there is an IPT above 3 GPa. Therefore, we have obtained the EoS at two different pressure ranges: before the IPT (1atm to 3 GPa) and after the IPT (3 GPa to 8 GPa). In addition, since the results of the three runs do not present significant divergences, only one adjustment was made on all the experimental points. Both experimental and theoretical data are summarized in **Table 3** showing rather good agreement.

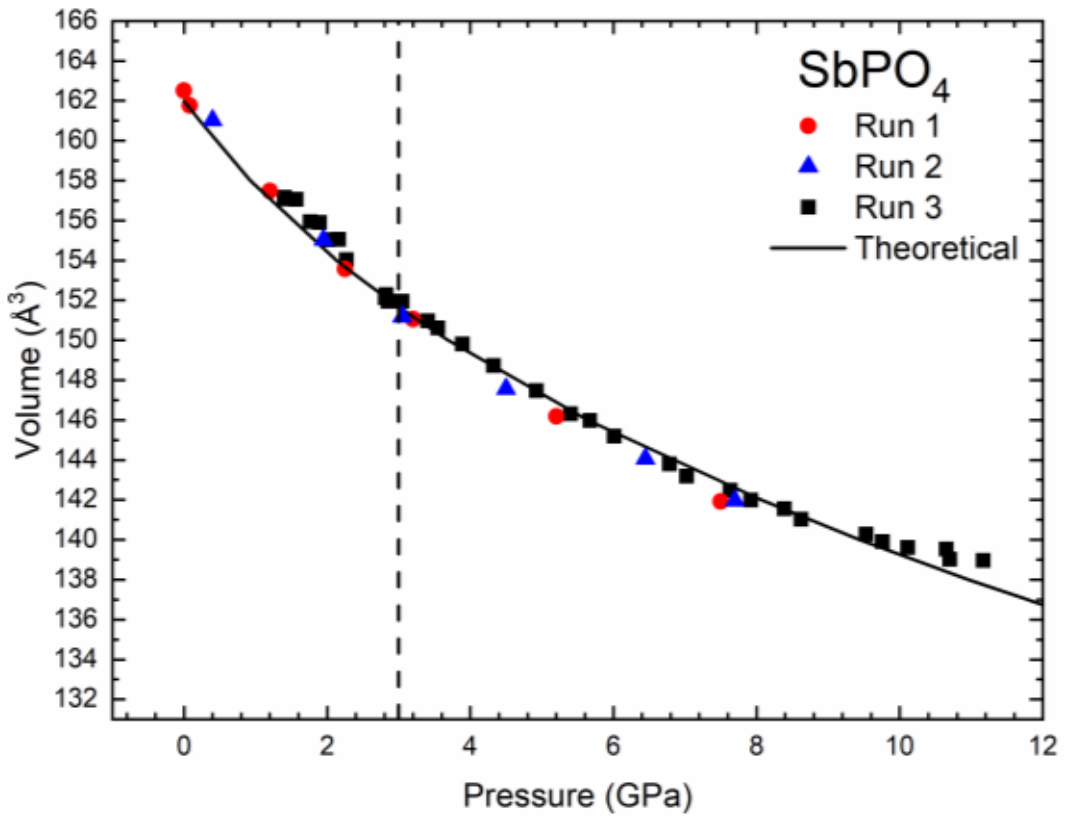

**Figure 3.** Unit-cell volume vs. pressure for $SbPO_4$. Symbols represent experimental data and solid line the theoretical data. The vertical dashed lines at 3 GPa indicate the pressure at which the IPT occurs.

**Table 3.** EoS parameters and axial compressibility ($\kappa_x = \frac{-1}{x}\frac{\partial x}{\partial P}$) of $SbPO_4$ before and after the IPT at 3 GPa. The variation $\frac{\partial x}{\partial P}$ was obtained using the modified Murnaghan equation of state $\Delta x_0 / x_0 = \left(1 + K'_{0x} P / K_{0x}\right)^{-\left(\frac{1}{3K'_0}\right)} - 1$, were $K_{0x}$ and $K'_{0x}$ are the bulk modulus and its pressure derivative of the *x*-axis (*x*=*a*, *b*, *c*, *β*) at atmospheric pressure.

| | | $V_0$ (Å$^3$) | $B_0$ (GPa) | $B'_0$ | $\kappa_a$ ($10^{-3}$ GPa$^{-1}$) | $\kappa_b$ ($10^{-3}$ GPa$^{-1}$) | $\kappa_c$ ($10^{-3}$ GPa$^{-1}$) | $\kappa_\beta$ ($10^{-3}$ GPa$^{-1}$) |
|---|---|---|---|---|---|---|---|---|
| Experimental | Up to 3 GPa | 162.6(6) | 36(3) | 6(2) | 14.6(1) | 5.2(4) | 8.5(6) | 3.8(5) |
| | From 3 GPa to 8 GPa | 160.3(8) | 45(2) | 7(3) | 9(1) | 4(1) | 9.7(6) | 1.4(2) |
| Theoretical | Up to 3 GPa | 163.0(2) | 32(1) | 9(3) | 17.93(4) | 2.7(2) | 14.25(3) | 5.4(1) |
| | From 3 GPa to 8 GPa | 160.6(4) | 45(2) | 6(2) | 12.8(2) | 2.9(6) | 11.6(1) | 4.6(15) |

As can be noticed, when fits are performed from 1 atm to 3 GPa, experimental and theoretical data yield $B_0$ values of 36(3) GPa and 32(1) GPa, respectively. However, after the IPT, both experimental and theoretical data yield a $B_0$ value of 45(2) GPa. The increase of $B_0$ is directly related to an increase of the structure rigidity after the IPT that is common in layered materials[59,66,67]. It is noteworthy of mentioning that, with a $B_0$ around 34 GPa, $SbPO_4$ is the most compressible phosphate[18]. Interestingly, the bulk modulus of $SbPO_4$ is almost half of that of barite-type compounds, such as $PbSO_4$ and $BaSO_4$[68–70] and is even smaller than the bulk modulus of the distorted barite-type structure of $SnSO_4$ and respective different layered phases[71]. This is noteworthy because the strong LEP of $Sn^{2+}$ of $SnSO_4$ leads to layered structures with a 3-fold coordinated Sn at the distorted barite-type *Pnma* structure and 3+1-fold coordinated Sn in the $P2_1/a$ phase above 0.2 GPa[71]. Consequently, we can safely conclude that $SbPO_4$ is not only the most compressible phosphate but also one of the most compressible $ABO_4$ compounds.

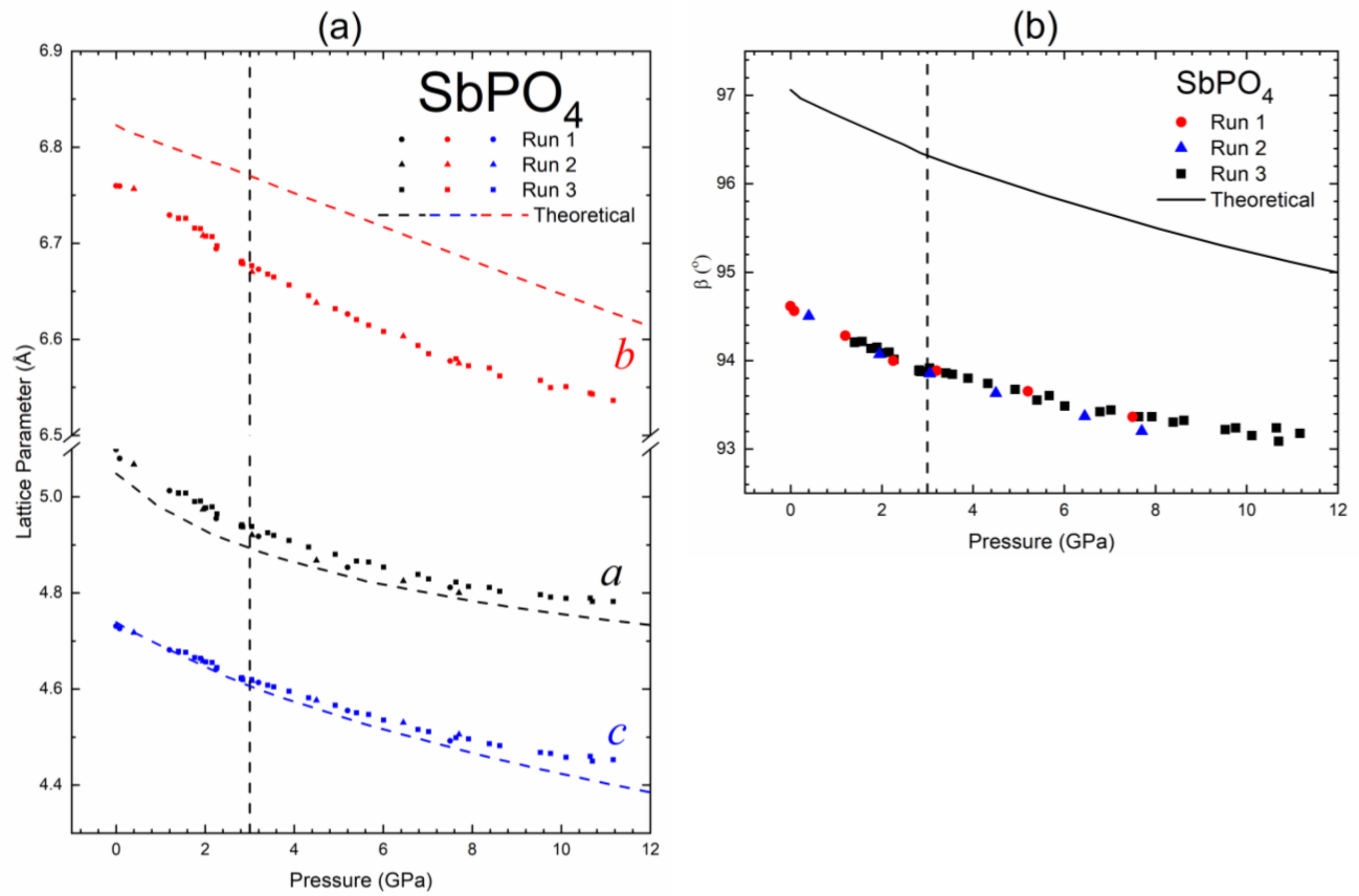

**Figure 4.** Experimental (symbols) and theoretical (lines) pressure dependence of (a) the lattice parameters *a*, *b* and *c* and (b) *β* angle. The vertical dashed lines at 3 GPa indicate the pressure at which the IPT occurs.

The pressure dependence of the experimental and theoretical lattice parameters *a*, *b*, *c* and *β* of monoclinic $SbPO_4$ is shown in **Figure 4**. The experimental unit-cell parameters as a function of pressure are represented only for P < 11.2 GPa. At higher pressures, phase coexistence does not allow to obtain them with reasonable accuracy. The axial compressibility, defined as $\kappa_x = -\frac{1}{x}\frac{\partial x}{\partial P}$ ($x$ = $a$, $b$, $c$, $\beta$), obtained from a modified Murnaghan EoS fit to the experimental data[65] is reported in **Table 3** and is in good agreement with our theoretical results. As previously, the adjustment was performed for two different pressure ranges: before the IPT (1 atm to 3 GPa) and after the IPT (3 GPa to 8 GPa). As expected for this layered material, the *a*-axis (direction perpendicular to the layers) presents the largest compressibility due to the high compressibility of the Sb LEP and the weak inter-layer Sb-O distances, and the *b*-axis evidenced the lowest value due to the small compressibility of the Sb-O3 and P-O3 bonds mainly directed along this axis. The parameter *a* and *c* present a significant decrease in the pressure coefficient at pressures higher than 3 GPa, thus supporting the hypothesis of an IPT around this pressure value. The β angle also presents a smooth decrease with pressure and, although our theoretical values present a discrepancy of ~2º in absolute value with respect to the experimental values, a similar evolution of the experimental and theoretical data with increasing pressure is evidenced. This result indicates that our theoretical data provides a correct description of the evolution of the lattice parameters and β angle of the monoclinic structure of $SbPO_4$ under compression.

Since $SbPO_4$ is a monoclinic material, we have calculated and diagonalized the experimental and theoretical isothermal compressibility tensor, $\beta_{ij}$, in order to evaluate the magnitudes and directions of the principal axes of compressibility[17]. The tensor has been calculated using the linear Lagrangian approximation (LLA)[72] and the infinitesimal Lagrangian approximation (ILA)[73]. For the LLA, a linear fit of the unit-cell parameters was carried out between the pressure range 0-5 GPa. **Table 4** summarizes the $\beta_{ij}$ coefficients of the isothermal compressibility tensor at zero pressure. It can be observed that there is a qualitative agreement between experimental and calculated data. In addition, the $\beta_{ij}$ coefficients obtained with LLA and ILA are also similar. Notice that $\beta_{22}$ is much smaller than $\beta_{11}$ and $\beta_{33}$ indicating that the *b*-axis is the less compressible axis of $SbPO_4$. On the other hand, from the experiments we obtain that $\beta_{11} > \beta_{33}$, but the opposite result is obtained from calculations. This is caused by the differences on the compressibility of the *a*-axis (*c*-axis) which is slightly underestimated (overestimated) by calculations.

Considering the eigenvalues obtained from experiments using LLA we obtain that the maximum, intermediate and minimum compressibilities are 10.22(6), 6.18(4), and 4.18(2) in units of $10^{-3}$ $GPa^{-1}$. Similar results are obtained from other approximations as can be seen in **Table 4**. These values are considerably larger than in $BiPO_4$ and $BiSbO_4$ [17,74] which is consistent with the layered structure of $SbPO_4$. The inverse trace of the compressibility tensor, expected to be equal to the bulk modulus, is 48 GPa, which agrees with the result obtained from the BM-EoS.

**Table 4.** Isothermal compressibility tensor coefficients, $\beta_{ij}$, and their eigenvalues, $\lambda_i$, and eigenvectors, $ev_i$, for $SbPO_4$ at room pressure. The results are given using the linear Lagrangian and the infinitesimal Lagrangian methods with data from our experiments and our *ab initio* simulations.

| Method | Linear Lagrangian | | Infinitesimal Lagrangian | |
|---|---|---|---|---|
| | Experiment | Theory | Experiment | Theory |
| $\beta_{11}$ ($10^{-3}$ $GPa^{-1}$) | 8.99(4) | 8.02 | 10.71(5) | 8.43 |
| $\beta_{22}$ ($10^{-3}$ $GPa^{-1}$) | 4.18(2) | 3.76 | 4.96(2) | 3.85 |
| $\beta_{33}$ ($10^{-3}$ $GPa^{-1}$) | 7.41(4) | 8.62 | 8.75(4) | 8.84 |
| $\beta_{13}$ ($10^{-3}$ $GPa^{-1}$) | -1.86(1) | -2.12 | -2.17(1) | -2.31 |
| $\lambda_1$ ($10^{-3}$ $GPa^{-1}$) | 10.22(6) | 10.47 | 12.11(6) | 10.96 |
| $ev_1$ ($\lambda_1$) | (0.834, 0, -0.557) | (0.655, 0, -0.755) | (0.840, 0, -0.542) | (0.675, 0, -0.738) |
| $\lambda_2$ ($10^{-3}$ $GPa^{-1}$) | 4.18(2) | 3.76 | 4.96(2) | 3.85 |
| $ev_2$ ($\lambda_2$) | (0, 1, 0) | (0, 1, 0) | (0, 1, 0) | (0, 1, 0) |
| $\lambda_3$ ($10^{-3}$ $GPa^{-1}$) | 6.18(4) | 6.17 | 7.35(4) | 6.33 |
| $ev_3$ ($\lambda_3$) | (-0.557, 0, 0.834) | (0.755, 0, 0.655) | (-0.542, 0, 0.840) | (0.738, 0, 0.675) |
| $\Psi$ (°) | 123(4) | 139 | 123(4) | 137 |

The eigenvalues and eigenvectors computed for the isothermal compressibility tensor are also reported in **Table 4**. Considering the eigenvector $ev_2$, the minor compression direction is along the *b*-axis. On the other hand, the major compression direction occurs along the (0 1 0) plane at the given angle $\Psi$ (see **Table 4**) to the ***c***-axis (from *c* to *a*). The direction of maximum compressibility, considering the value of the $\beta$ angle is at 30(4)º (42º) to the *a*-axis for the case of our experiments (calculations). The direction of intermediate compressibility is also at the same plane, but it is perpendicular to the direction of maximum compressibility. Graphically, the directions of both maximum and intermediate compressibility at room pressure can be observed in **Figure S7** in SI.

Using the results of the experimental and theoretical lattice parameters, we have plotted the pressure dependence of the *c/a*, *c/b* and *a/b* axial ratios (**Figure 5a-c**). The results evidence a good agreement of the trends of the experimental and theoretical axial ratios with pressure. The *c/b* parameter only presents a smooth decrease at HP. On the other hand, the *c/a* and *b/a* ratios show a significant increase up to ~2 GPa and ~3 GPa, respectively. At higher pressures, the *c/a* parameter decreases and the *b/a* trends to stabilize. In **Figure 5d** we plot the normalized pressure vs eulerian strain plot. As can be observed, both experimental and theoretical F-f data show a change in slope at Eulerian strain values at ~0.02 (~3 GPa) and 0.009 (~2 GPa), respectively. The strong change observed in the slopes of the *c/a* and *b/a* axial ratios and the F-f plot above 3 GPa seem to suggest an IPT around that pressure range, which will be further discussed.

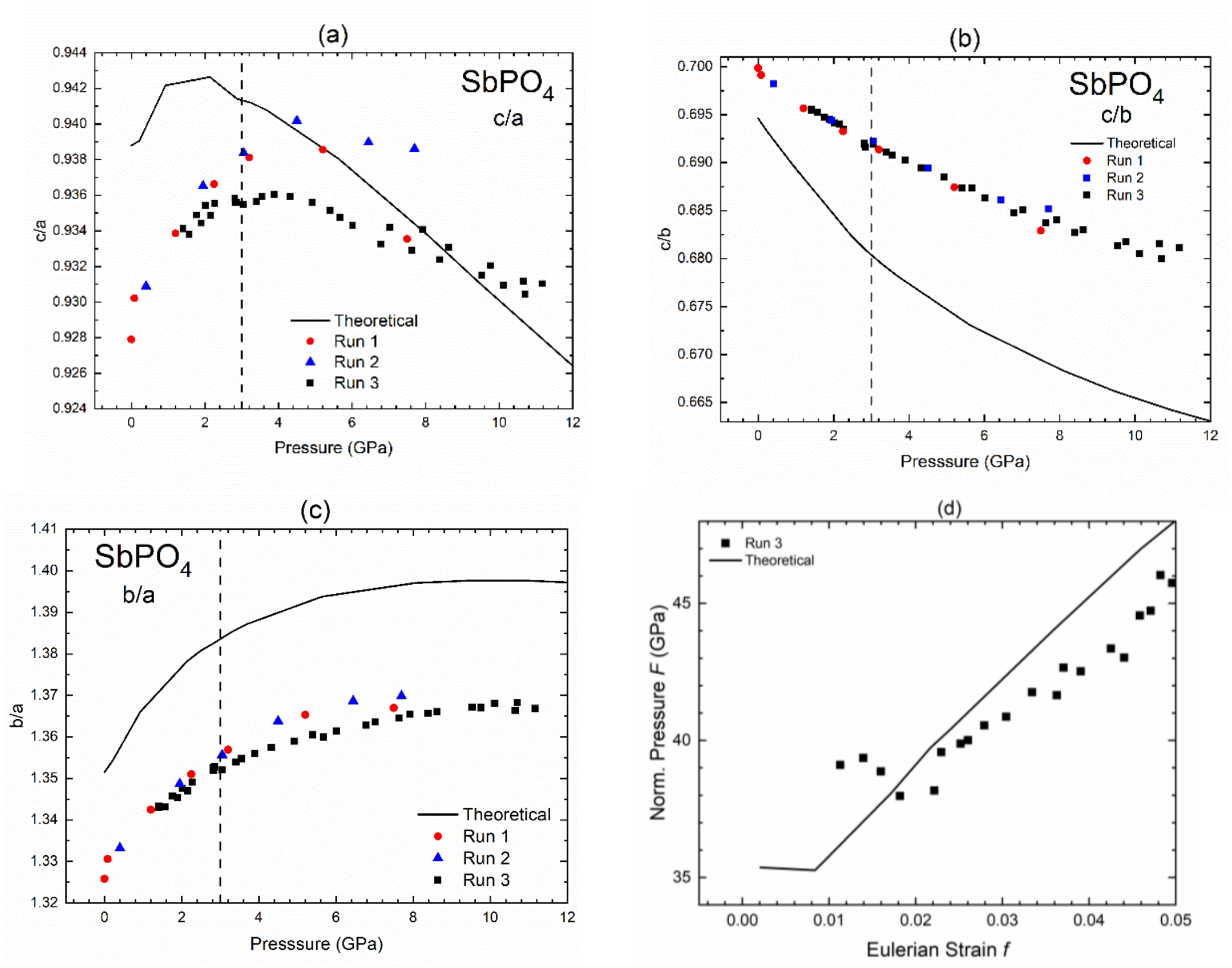

**Figure 5 -** Experimental (symbols) and theoretical (lines) pressure dependence of (a) *c/a* (b) *c/b* and (c) *b/a* ratio, and (d) normalized pressure vs eulerian strain plot. The vertical dashed lines at 3 GPa indicate the pressure at which the IPT occurs as suggested by the change of the pressure coefficients of the axial ratios. Only experimental data of run 3 have been used for the normalized pressure vs eulerian strain plot due to the extreme sensitivity of this plot to data dispersion.

Considering the good correlation between our experimental and theoretical results for monoclinic $SbPO_4$, we can use the theoretical results to extract additional information that is not available through the LeBail fit, such as the evolution of the free atomic positions, bond lengths and polyhedron distortion at HP. In **Figure 6**, we can observe the pressure dependence of the theoretical Sb-O and P-O bond lengths. As can be noticed in **Figure 6(a)**, the shortest Sb-O1 bond length shows no significant change with pressure, but the shortest Sb-O2 and Sb-O3 bond lengths (see solid lines in **Figure 1(b)**) tend to converge to the same value as the pressure increases. In this context, it is worth mentioning the increase of the Sb-O2 bond length between 0 and 3 GPa and its change of slope above 3 GPa. Similar changes of slope close to 3 GPa can also be observed at other Sb-O distances. As regards the largest Sb-O lengths (marked with * in **Figure 6(a)**), which correspond to the two inter-layer Sb-O3 distances and the two dashed lines shown in **Figure 1(b)**, these show a considerable decrease below 3 GPa. Above this pressure value, this tendency decreases, but is still reminiscent. Similarly, all P-O bond lengths (**Figure 6(b)**) decrease with pressure, except the P-O3 bond that remains almost constant below 3 GPa and decreases above this pressure.

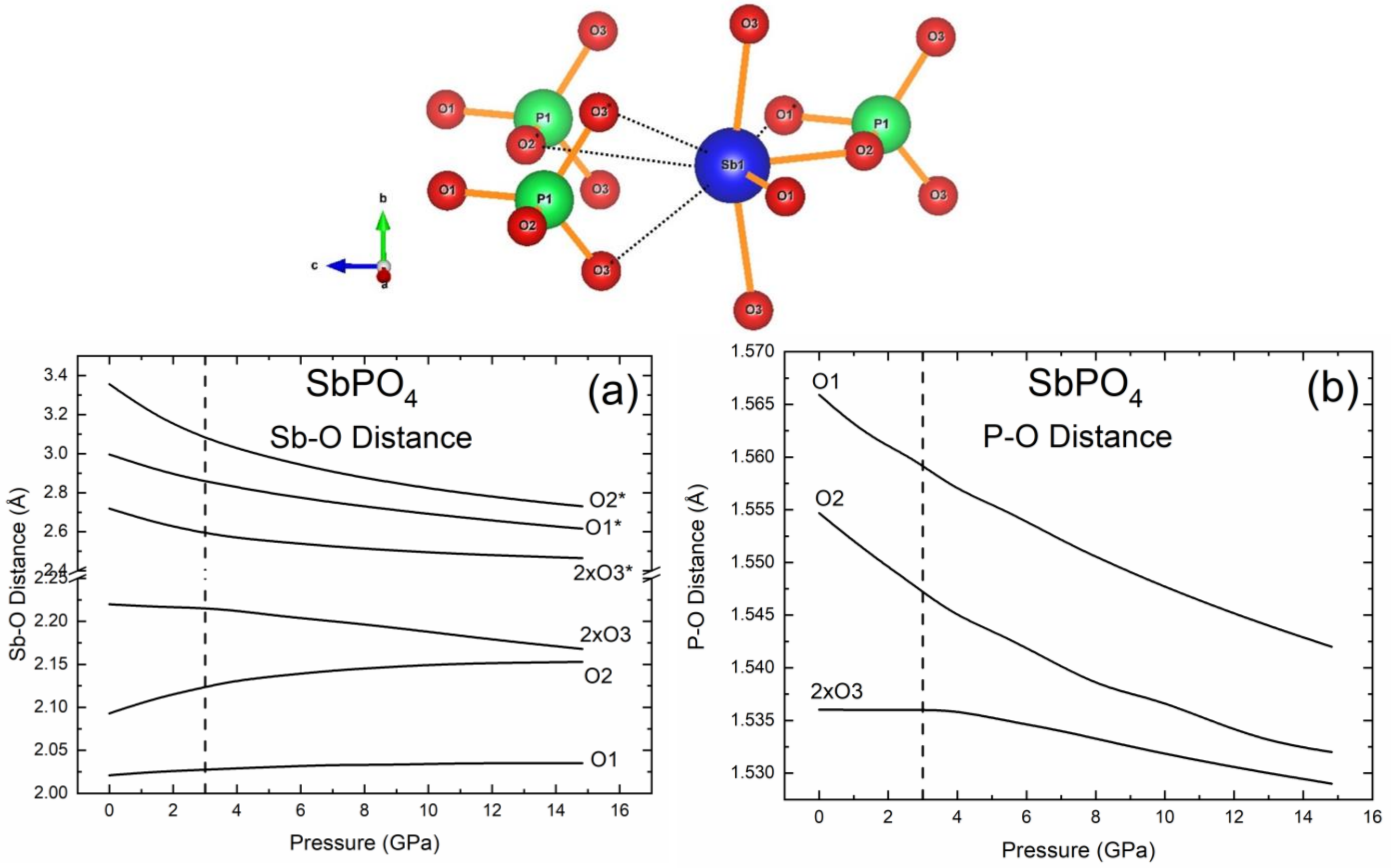

**Figure 6.** Evolution of cation-anion distances of $SbPO_4$ with pressure increase. The * indicate the oxygen distances of adjacent layer. The results were obtained by theoretical calculations. The vertical dashed lines at 3 GPa indicate the pressure at which the IPT occurs as suggested by the change of most pressure coefficients of the interatomic distances.

As already commented, changes of the slopes for the many bond lengths of the monoclinic $SbPO_4$ are observed around 3-4 GPa, especially for distances related to the O3 atom; i.e. the external O atoms of the layers, while smaller changes are associated to O1 and O2 atoms; and the internal O atoms of the layers (see **Figures 1(a)** and **(b)**). To trace the origin of those changes we have plotted in **Figure S8** of SI the pressure dependence of the Wyckoff sites of monoclinic $SbPO_4$. In order to assure the good agreement between our theoretical and experimental data, the experimental values obtained by Rietveld refinement at room pressure were also included in **Figure S8**. As can be observed, the *z* value of all sites tends to decrease with pressure, except for Sb. It is also possible to observe that the evolution of all positions presents a minor change of the slope around 3 GPa; however, the largest variation of the slope is observed for the *x* position of both O2 and O3 and the *y* position of O3 (**Figures S8(d) and**

**S8(e))**. These trend variations are indicative of a pressure-induced IPT close to 3 GPa, as previously commented. Further discussion about the IPT will be provided when we discuss the behavior of the electron topology at HP.

In order to find the origin of the new peaks above 8.4 GPa, we provide an indexation of the XRD pattern above that pressure assuming the possibility of a phase coexistence between the LP phase and a new HP phase or assuming a single HP phase. To search for possible HP structures of $ABO_4$ compounds, we resorted to the Bastide's diagram[20,32] by taking into account the position of $SbPO_4$ in that diagram ($r_{Sb}/r_O$ = 0.563, $r_P/r_O$ = 0.126). We have plotted a renewed form of the Bastide's diagram in **Figure 7** highlighting the location of many compounds containing cations with LEPs, such as $As^{3+}$, $Sb^{3+}$ and $Bi^{3+}$, which have been positioned in respective diagram for the first time. We must stress that there are many compounds with $As^{5+}$ in the diagram, with $As^{5+}$ behaving similarly to $P^{5+}$, but there is only one compound of $As^{3+}$ (in red), which is precisely $AsPO_4$ with $P^{5+}$. We must also stress that there are many compounds with $P^{5+}$ in the diagram, but no compound with $P^{3+}$ is mentioned.

According to the north-east rule in Bastide's diagram, $SbPO_4$ should crystallize in the orthorhombic $CrVO_4$-type structure with [6-4] coordination for the [*A*-*B*] cations. Moreover, this compound should transform into a zircon or scheelite phase under compression (see black arrow in **Figure 7**). However, neither these structures allow us to explain the new peaks observed in **Figure S6**. Since $BiPO_4$-III transforms into the monazite structure above 0.8 GPa[12], we have also tried the comparison of the peaks with the monazite structure. However, the position of the new diffraction peaks cannot be explained with a possible PT to this structure either, despite the monazite structure being energetically more favorable than the monoclinic structure at HP.

Since Sb is 4-fold coordinated for the monoclinic structure instead of 6-fold coordinated, as expected from the $Sb^{3+}$ ionic radius, we have considered that the real position of $SbPO_4$ could be that of $AsPO_4$, which is predicted to have 4-fold coordination for As, despite of existing a real 3-fold coordination of As at room pressure[75]. In such a case, monoclinic $SbPO_4$ could transform under pressure into the $CrVO_4$ or wolframite structures (see red arrow in **Figure 7**); however, the positions of the new diffraction peaks cannot be explained with a possible PT to these structures either. Other candidate structures for the HP phase of $SbPO_4$ were also considered: $TiPO_4$, $BaSO_4$, $HgSO_4$, $AgMnO_4$, $BaWO_4$-II, $BiSbO_4$ and the different structures of $SnSO_4$. Note that the $Sn^{2+}$ of $SnSO_4$ also features a strong LEP that leads to the crystallization of $SnSO_4$ in a distorted barite structure. In this context, several compounds featuring other cations with strong LEPs, such as $Sn^{2+}$ and $Pb^{2+}$, are also shown in **Figure 7**. None of these structures (by themselves or coexisting with the LP phase) enables us to clarify all the Bragg peaks observed for $SbPO_4$ above 8.4 GPa.

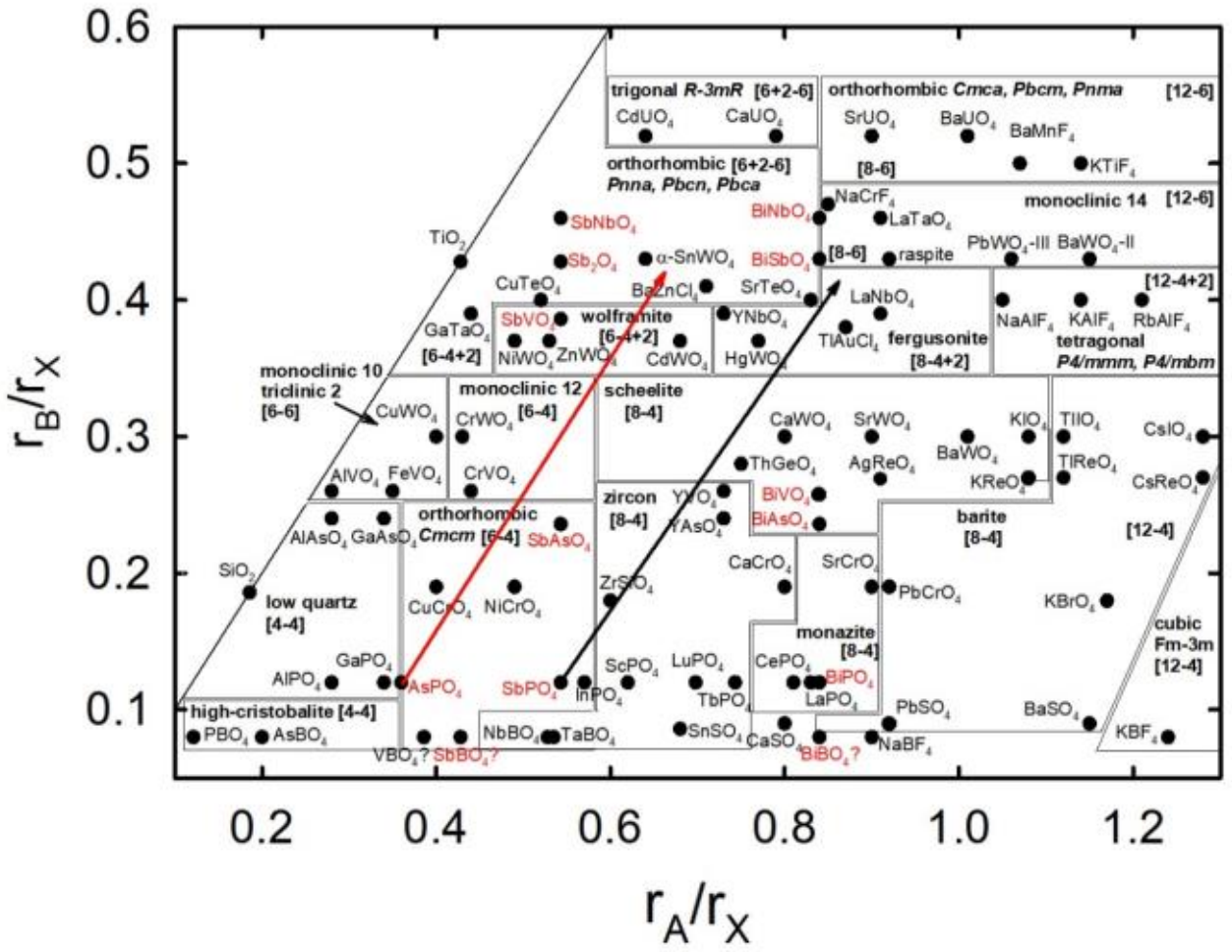

**Figure 7.** Bastide's diagram for *AB*$O_4$ compounds including the new family of borates and all known compounds containing As, Sb and Bi with valence $3^+$.

Finally, we found a possible solution by considering the coexistence of the LP monoclinic phase of $SbPO_4$ with a respective triclinic distortion. Such coexistence has been observed in other monoclinic oxides at HP[71,76]. In our case, we have built the candidate triclinic structure, which belongs to the *P*-1 (No. 2) space group, by using the group-subgroup relationships between space groups n° 2 and 11. By considering the coexistence of the LP monoclinic structure and the HP triclinic structure, we have been able to clarify the diffraction patterns measured above 8.4 GPa. Our XRD patterns suggest that the LP phase is the dominant phase up to 11.2 GPa, being the HP phase the dominant phase above this pressure value. In **Figure 8**, we show the result of the profile matching of the XRD pattern at 15.2 GPa by fixing the atomic coordinates of both monoclinic and triclinic phases to the calculated ones at that pressure range. The small residual of the fit supports the hypothesis that the triclinic HP phase is a distortion of the monoclinic LP phase. The R-values of the fit shown in **Figure 8** are $R_p = 6.5\%$ and $R_{WP} = 8.5\%$. At 15.2 GPa, the unit-cell parameters of the monoclinic LP phase are: $a$ = 4.764(8) Å, $b$ = 6.502(9) Å, $c$ = 4.423(8) Å, and $\beta$ = 92.93(9)°, with $V_0$= 136.8(7) $Å^3$; while the unit-cell parameters of the triclinic HP phase are: $a$ = 4.704(8) Å, $b$ = 6.443(9) Å, $c$ = 4.531(8) Å, $\alpha$ = 92.85(9)°, $\beta$ = 92.93(9)°, $\gamma$ = 92.47(9)°, with $V_0$= 136.8(8) $Å^3$. As observed, the volumes of both LP and HP phases at this pressure range are similar, being the experimental values of the triclinic HP phase close to those calculated at 15.0 GPa (see **Table S2)**. This result and the group-subgroup relationship between both structures suggest that the PT could be a very weak first order transformation, as suggested by the coexistence of both monoclinic and triclinic structures at HP and the reversibility of the XRD pattern at room pressure previously mentioned.

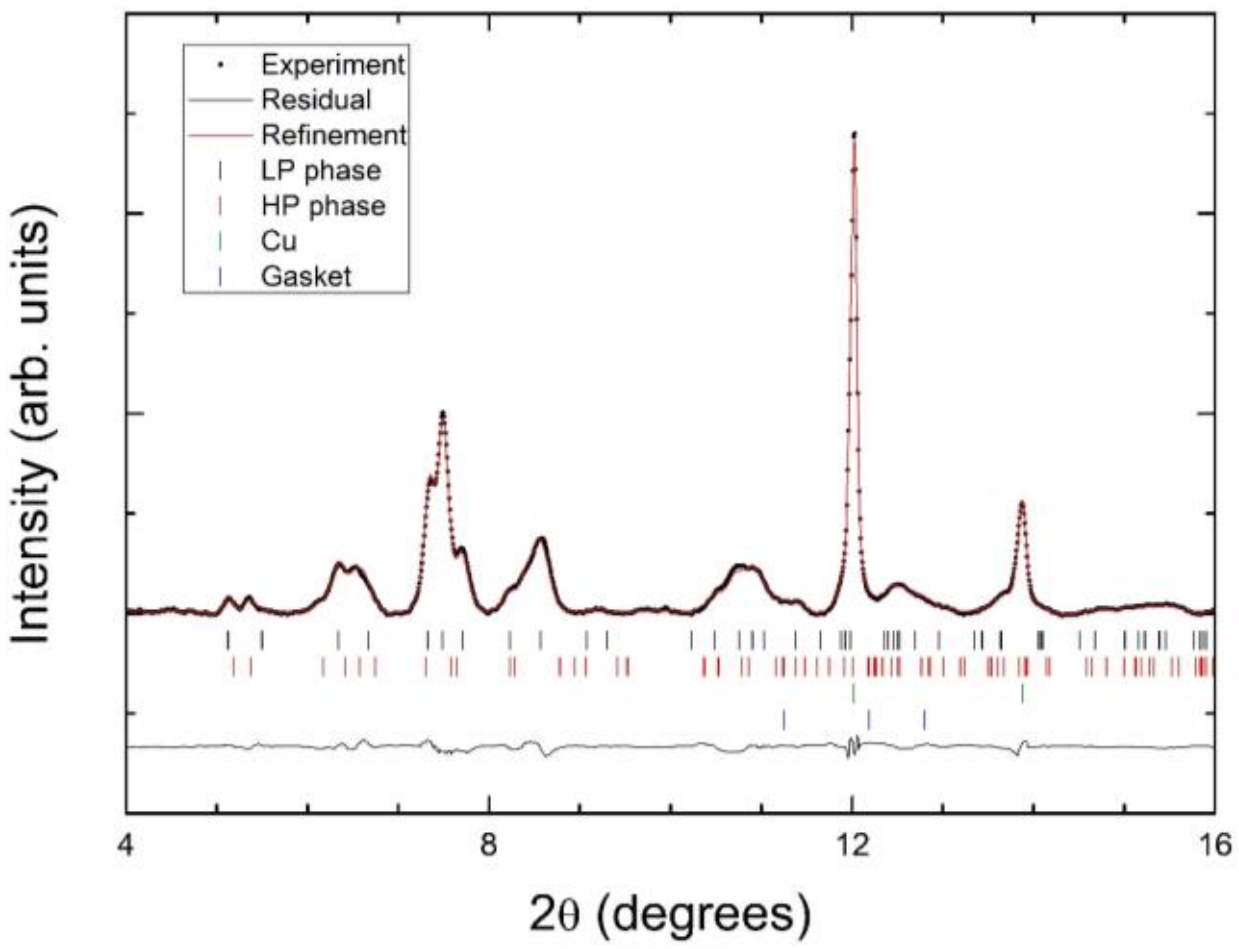

**Figure 8.** Angle-dispersive XRD of $SbPO_4$ measured at 15.2 GPa at room temperature. Experiments, refinements and residuals are shown. The ticks indicate the position of the peaks of different phases.

Support for the monoclinic-to-triclinic PT stems from results obtained from enthalpy vs pressure calculations for both phases which evidence a PT at pressure values of ~10 GPa (see **Figure S9** in SI). Nevertheless, there is a possibility that the PT (structural distortion) could be triggered by non-hydrostatic effects[77]. This fact deserves to be studied in the future by further experiments using a selection of different pressure-transmitting media. It must also be stressed that we have found other triclinic structures with lower energy values than the triclinic structure proposed as HP phase for $SbPO_4$. These other triclinic structures are the HP structure of $SnSO_4$ observed above 13.5 GPa[71]. However, our XRD patterns at 15.2 GPa cannot be fitted to such structures. Note that these triclinic structures of $SnSO_4$ feature a double unit-cell with twice the number of atoms per unit-cell that in our triclinic phase. This would lead to a much larger number of vibrational modes than those observed above 9 GPa, so we may safely disregard such structures, as we will show in the following section.

Finally, we used the VESTA software[78] in order to evaluate the effective coordination number (ECoN)[79,80] of the Sb and the distortion index[81] of Sb polyhedron in the $P2_1/m$ and $P$-1 phases of $SbPO_4$ at different pressures (**Figure 9**). The ECoN scheme has been recently discussed by Gunka and Zachara and shown to be very helpful in discussing the coordination of cations with LEP activity[82]. For this purpose, the pressure dependence of the Sb-O interatomic distances of the eight next-neighbor O closest to Sb have been obtained from the calculated structures at different pressures. As can be observed in **Figure 8(a)**, the Sb ECoN of the LP phase of $SbPO_4$ (3.83 at 0 GPa) is consistent with the 4-fold coordination of Sb at room pressure. The ECoN increases steadily with pressure in the LP phase and reaches 4.78 at 10.97 GPa thus pointing to a 4+2-fold coordination of Sb at this pressure since there are two additional Sb-O3 inter-layer distances with the same length. Above this pressure, the HP phase becomes more stable and the evolution of the ECoN of Sb of the HP phase presents the same growth rate as that of the LP phase, thus reaching an ECoN of 5.19 at 18.5 GPa. We will see later that this value is consistent with a 4+2+1-fold coordination for Sb coordination for this pressure range. The increase of ECoN with pressure is followed by the decrease of the distortion index (**Figure 9(b)**) that, above 4.7 GPa presents a decrease of the distortion rate and remains constant in the HP phase. Moreover, the increase of the Sb coordination in the monoclinic phase from 4 to 4+2 can be related to the strong decrease of the Sb eccentricity of the $SbO_6$ polyhedron between 1 atm and (see **Figure S10**).

Finally, we must stress that the ECoN value of Sb in $SbPO_4$ at 18.5 GPa is close to that of Bi for $BiPO_4$-III at room pressure (5.16); therefore, we can conclude that around 18 GPa the $SbPO_4$ compound behaves as $BiPO_4$-III at room pressure[19]. In other words, pressure promotes the approach of the layers in $SbPO_4$, thus favoring the bond between the $Sb^{3+}$ of one layer and the $O^{2-}$ atoms of the adjacent layer, therefore converting the 2D-type structure of $SbPO_4$ at room pressure into a 3D-type structure that reaches a similar coordination to that of $BiPO_4$-III at pressures close to 18 GPa.

We conclude by mentioning that the proposed pressure-induced IPT at 3 GPa and monoclinic-triclinic PT above 9 GPa does not involve a change of the coordination of P, although a considerable increase of the coordination of Sb from 4 at room pressure to 4+2 above 3 GPa is observed; moreover respective coordination increases to 4+2+2 above 9 GPa (see **Figure 1(c)**). In fact, we may consider the IPT at 3 GPa as being the onset of a 2D-to-3D PT due to the increase of coordination to 4+2 caused by two new Sb-O3 inter-layer bonds. This can be seen as a gradual distortion of the crystal structure favored by the presence of a LEP, which gives a large flexibility to the crystal structure to accommodate stresses/strain. It is an analogous phenomenon to what we have seen in $CuWO_4$ [83], where there is a pressure-induced structural distortion in order to preserve the Jahn-Teller distortion of the compound. On the other hand, the coordination of Bi of the $BiPO_4$-III structure at 0 GPa is practically 6, which explains the reason why this material does not form layers at room pressure, unlike $SbPO_4$, and why $BiPO_4$-III varies directly to coordination 6 + 2 at HP (in the monazite phase)[19].

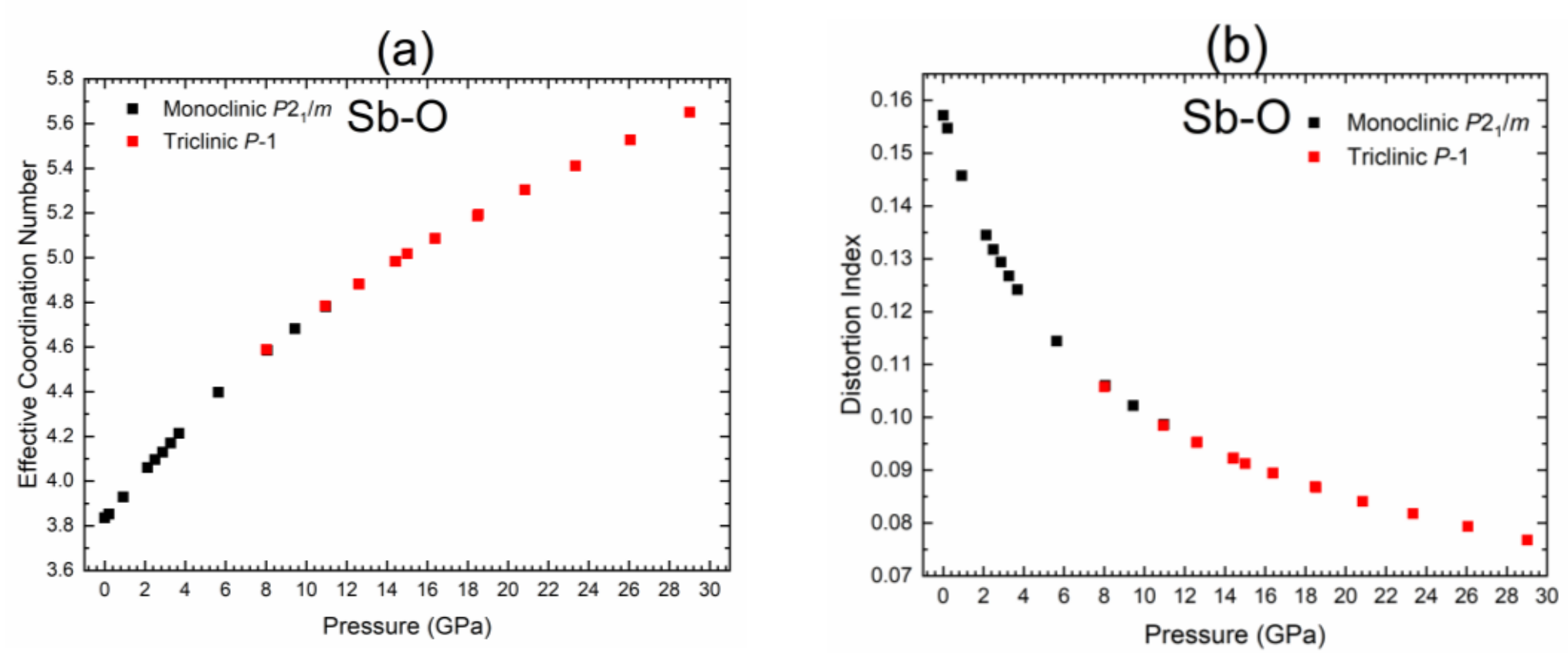

**Figure 9.** Pressure dependence of the (a) Effective Coordination Number (ECoN) and (b) distortion index of the $Sb^{3+}$ of the monoclinic (black squares) and triclinic (red squares) structures of $SbPO_4$ as obtained from our theoretical calculations using the VESTA software.

## 4.3. Vibrational properties under compression

Raman scattering (RS) spectra at selected pressures up to 24.5 GPa are presented in **Figure S11**. Once the sample is inside the pressure cell, it is possible to observe some peaks that probably are not related to the $SbPO_4$ sample (see blue arrows in **Figure S11)** since these do not appear on the RS spectra at room pressure either before or after the HP cycle (see bottom and top RS spectra in **Figure S11)**. These peaks could be due to some unintentionally impurity loaded on the DAC. The pressure dependence of these peaks is plotted as blue symbols in **Figure 10** and some of them can be observed up to the maximum pressure of our RS experiment.

As regards to the peaks that may be considered as first-order modes of $SbPO_4$, some of these begin to widen and lose intensity and other new peaks start to rise above 7.7 GPa (see red arrows in **Figures S11(a)**, **(b)** and **(c)**), thus giving support to the existence of a PT above this pressure range. In particular, the peak that rises at 12.8 GPa around 134.7 $cm^{-1}$ and at 24.5 GPa around 160.4 $cm^{-1}$ (**Figure S11(a)**) becomes the most intense peak of the RS spectrum of the HP phase. Other seven less intense new peaks can be observed between the pressure range of 7.7 and 16.2 GPa. Notably, the peaks initially observed at 355 $cm^{-1}$ (the strongest one of the LP phase) and at 888 $cm^{-1}$ (probable second-order mode of the LP phase) progressively disappear with increasing pressure, thus indicating that the PT seems to be completed around 20 GPa. This result could explain why our XRD measurements up to 15 GPa cannot clearly resolve the HP phase since this phase is not completely developed at this pressure range. Note also that the region, which presents less changes of the Raman spectrum is the high-frequency region related to the stretching P-O vibrations of the $PO_4$ unit. This means that the HP phase is most likely to be a phase with tetrahedral coordination of P, in good agreement with the proposed triclinic HP phase and with the higher pressure phase at which the P coordination has been observed to increase on other phosphates[30,84].

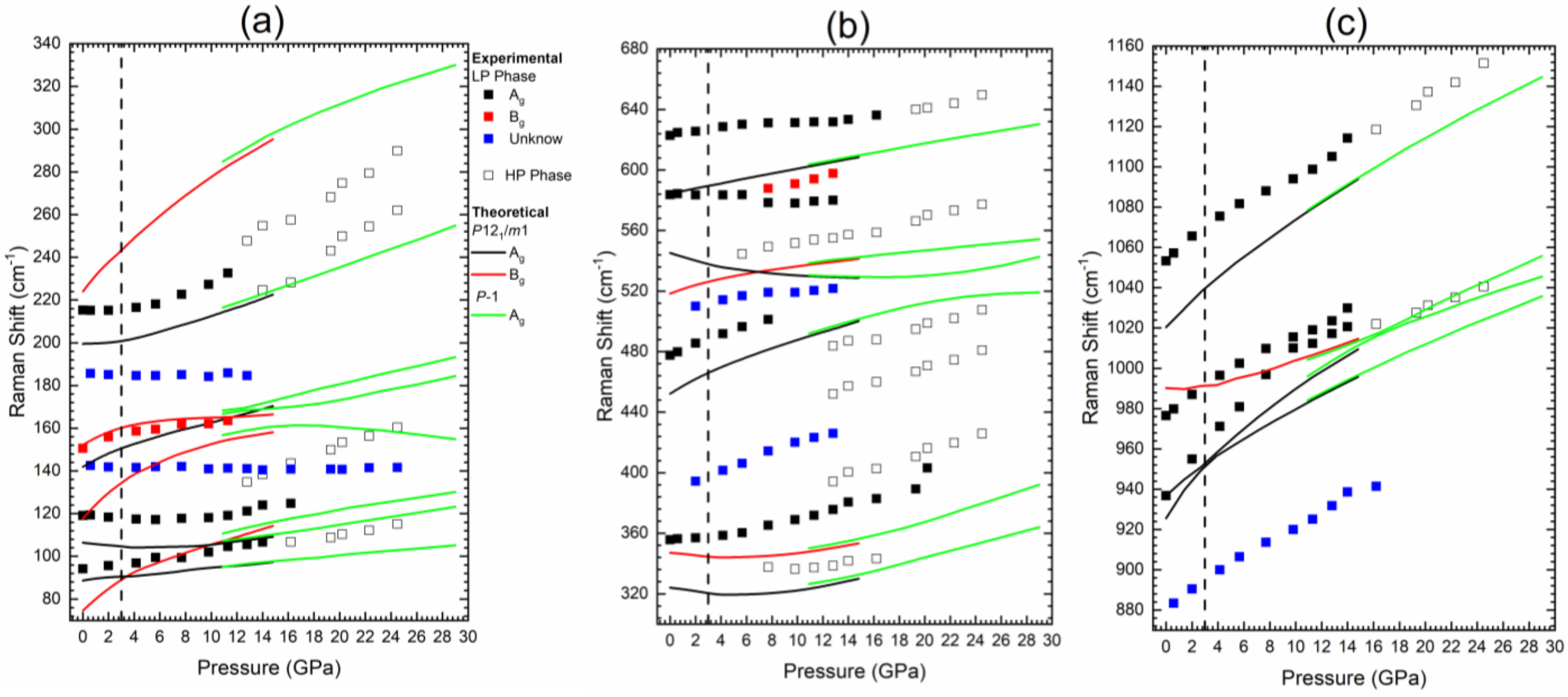

**Figure 10.** Experimental (Black - $A_g$ and Red - $B_g$ symbols) and theoretical (lines) pressure dependence of the Raman-active modes of $SbPO_4$: (a) from 75 to 300 $cm^{-1}$; (b) from 300 to 750 $cm^{-1}$, (c) from 870 to 1190 $cm^{-1}$. Blue symbols represent peaks that are not related to the $SbPO_4$. Lines with different colors represent Raman-active modes of different symmetries. Open symbols represent the new peaks not related to the initial phase. Blue symbols represent the peaks that is not related to the initial $SbPO_4$ phase. The vertical dashed lines at 3 GPa indicate the pressure at which the IPT occurs as suggested by the change of many frequency pressure coefficients.

**Figure 10** presents the dependence of the experimental and theoretical frequencies of the Raman peaks of $SbPO_4$ at HP, which is also summarized in **Table 2**. For the sake of completeness, we also plotted the dependence of the theoretical IR-active modes at HP in **Figure S12** in SI, which data are summarized in **Table S3** in SI. Comparing the evolution of the theoretical and experimental results at HP, we can note that results obtained from *ab-initio* calculations underestimate the frequencies of all Raman-active modes. This underestimation (typically within 3-5%) is especially evident in the medium- and high-frequency regions, where frequency values differ up to 30 $cm^{-1}$. However, comparing the pressure evolution of both data, we can tentatively assign the symmetry irreducible representations of some experimental Raman-active modes with the aid of theoretical calculations (see **Table 2** and **Figure 10**). For this purpose, we have calculated the pressure coefficients of the Raman peaks up to 3 GPa (**Table 2**) due to the IPT observed above 3 GPa. Curiously, all experimentally observed peaks at room pressure can be associated to the $A_g$ modes, except for the peak located at 151 $cm^{-1}$, which we attribute to the $B_g$ mode at 151 $cm^{-1}$. Finally, it must be mentioned that the signature of the experimental broad peak initially observed at 107 $cm^{-1}$ is not clear since it was observed only at 1 atm outside the DAC (before and after the pressure cycle).

As can be observed in **Figure 10**, many vibrational modes present a change in the pressure coefficient between 3 and 6 GPa, reinforcing the idea of the existence of a pressure-induced IPT around 3 GPa. In particular, experimental Raman-active modes Ag(T) (near 215 $cm^{-1}$) and Ag(R) (near 356 $cm^{-1}$) as well as a number of theoretical Raman-active modes (at 75, 106, 118, 152, 200, 324, 347, and 990 $cm^{-1}$ at 0 GPa in **Table 2**) show a change of slope close to 3 GPa in **Figure 9**. Moreover, all the vibrational modes of $SbPO_4$ that show a negative pressure coefficient at 0 GPa change to a positive pressure coefficient above 3 GPa. This result is in good agreement with the pressure-induced 2D-to-3D phase transition that takes place in layered $SbPO_4$ above 3 GPa upon increasing Sb coordination from 4 to 4+2-fold. On the other hand, the non-linear behavior of the theoretical vibrational modes located at 926 and 937 $cm^{-1}$ at room pressure, is the result of an anticrossing of these two $A_g$ modes, which is reproduced by the experimental results at a slightly higher-pressure value (~12 GPa – **Figure 10(c)**). A change of pressure coefficient around 3 GPa can also be observed for many theoretical IR-active modes (**Figure S12**), where a couple of anticrossings seem also to be observed for the $B_u$ peaks 183 and 207 $cm^{-1}$ (**Figure S12(a)**) and at 930 and 937 $cm^{-1}$ (**Figure S12(c)**), respectively. The change of the pressure coefficient of the Raman-active and IR-active modes near 3 GPa can be related to the approximation of the atomic layers that begin to interact more strongly and lead to the increase of Sb coordination. Note that the compression of the LEP is much larger than that of other bonds, thus leading to a large compression of the inter-layer distance below 3 GPa (compression is less pronounced at higher pressures).

At this point, we can discuss the pressure coefficients of the vibrational modes. It can be observed that the largest pressure coefficients correspond mostly to the P-O vibrations stretching located at the high-frequency region. In particular, the highest-pressure coefficient is that of the symmetric stretching $A_g$ mode and the respective IR analogue, the $B_u$ mode. A similar high response to pressure of the stretching P-O vibrations, and in particular of the symmetric stretching modes, has been found for other orthophosphates[19,20,26,85–88]. Large pressure coefficients are also observed for the rotational modes of the $PO_4$ unit (theoretical $A_u$ and $B_g$ modes at 220 and 224 $cm^{-1}$, respectively). Again, this behavior has already been observed for other orthophosphates[19,20,26,85–88].

As regards to the rigid layer modes, the shear rigid layer modes positioned at 75 $cm^{-1}$ ($B_g$ mode) and at 89 $cm^{-1}$ ($A_g$ mode) have pressure coefficients of 4.5 and 0.5 $cm^{-1}$/GPa, respectively. On the other hand, the longitudinal rigid layer mode at 106 $cm^{-1}$ ($A_g$ mode) has a pressure coefficient of -0.6 $cm^{-1}$/GPa (see **Table 2**). For typical layered materials, with van der Waals interaction between the layers, such as GaSe and InSe, the longitudinal rigid layer mode has a larger pressure coefficient (above 3 $cm^{-1}$/GPa) when compared to that of the shear rigid layer mode (between 0.5 and 1.5 $cm^{-1}$/GPa, see discussion in **Refs. 59** and **89**). The situation of $SbPO_4$ is completely different to that of typical layered compounds but also different to that of BiTeBr and BiTeI with polar interactions between the layers[59]. On one hand, the lowest-frequency $A_g$ mode is a typical shear rigid layer mode (see **Figure S2**) and evidences a pressure coefficient below 1 $cm^{-1}$/GPa. On the other hand, the shear rigid layer $B_g$ mode shows an extraordinarily high-pressure coefficient. This can be explained taking into account the atomic vibrations of this latter mode (see **Figure S1**). It can be observed that the $B_g$ mode is not a pure shear mode because it involves mainly motion of the Sb atom with both Sb and O of the same sublayer vibrating out-of-phase. Therefore, this mode is a mixture of an asymmetric stretching (of Sb-O3 bonds) and bending of Sb-O1 and Sb-O2 bonds within the $SbO_4E$ unit, what justifies the high value of the pressure coefficient of this mode.

Finally, we want to highlight that the negative pressure coefficient for the longitudinal rigid layer mode is a characteristic feature of $SbPO_4$ not reported for any other layered compound to our knowledge. A positive pressure coefficient for this mode has been observed in all van der Waals-type layered compounds, i.e. InSe and GaSe and other related materials, and also for layered compounds with polar interlayer interaction, such as BiTeBr and BiTeI[59]. The positive value of this pressure coefficient for van der Waals-type compounds is related to the increase of the interlayer strength with increasing pressure. A closer look at the atomic vibrations of this mode shows that this mode is also a mixture of an asymmetric stretching (of Sb-O2 bond) and bending of Sb-O1 and Sb-O3 bonds within the $SbO_4E$ unit. Therefore, the negative pressure coefficient for this mode in $SbPO_4$ is most likely related to a decrease of the Sb-O2 bond strength which is in good agreement with the increase of the Sb-O2 bond distance between 0 and 3 GPa (see **Figure 6**). Note that the change of the pressure coefficients of many vibrational modes is also in agreement with the changes of the Sb-O distances observed in **Figure 6**, thus providing additional support to the occurrence of a second-order IPT for $SbPO_4$ around 3 GPa.

Several new Raman-active modes (**Figure S11**) observed above 12.8 GPa have been attributed to the HP phase. Group theoretical considerations for the proposed triclinic (*P*-1) HP phase yield 36 normal modes of vibration at **Γ**, which mechanical decomposition has the form[57]:

$$\mathbf{\Gamma} = 18\ A_g(R) + 15\ A_u(IR) + 3\ A_u$$

where $A_g$ are Raman-active (R) and $A_u$ are IR-active, except for the three acoustic modes. Therefore, there are eighteen Raman-active and fifteen IR-active modes. The eighteen Raman-active and fifteen IR-active theoretical modes have been plotted in **Figures 10** and **S13**, respectively. As observed in **Figure 10**, the theoretical Raman-active modes for the HP phase of $SbPO_4$ show similar frequencies and pressure coefficients to those of the LP phase. **Table 5** summarizes the frequencies and pressure coefficients of the experimental and theoretical modes of the HP triclinic phase of $SbPO_4$. Despite of not existing a very good agreement between the experimentally measured modes of the HP phase and the calculated ones, we have provided in **Table 5** a tentative assignment of the experimental modes to this triclinic phase. The theoretical frequencies and pressure coefficient of the IR-active modes of the proposed triclinic HP phase of $SbPO_4$ are also summarized in **Table S4** in SI. Regarding the relative disagreement between calculated and experimental triclinic Raman-active modes in **Table 5**, we think that it can be due to experimental problems of appearance of second-order modes instead of first-order modes of the triclinic phase or to theoretical problems regarding the simulation of the correct triclinic phase since the experimental triclinic phase could be slightly different to the simulated one. Regarding this point, we must note that simulation of triclinic phases is very challenging since energy minimization procedures can lead to local minima and not to absolute minima. This means that we have found a triclinic phase that is competitive with the monoclinic one at HP, but we cannot assure that this is the only triclinic competitive phase and therefore we cannot assure that the simulated one is exactly the experimental one.

In summary, our unpolarized HP-RS measurements of $SbPO_4$ exhibit most of the Raman-active modes of the monoclinic ($P2_1/m$) phase with $A_g$ symmetry, but very few modes with $B_g$ symmetry. The assignment of vibrational modes as internal or external of the $PO_4$ units has been provided and their pressure coefficients, especially those for rigid layer modes, properly discussed. HP-RS results support the occurrence of an IPT around 3 GPa and a PT above 8 GPa that complete respective formation around 20 GPa, in good agreement with the XRD measurements. Finally, the Raman-active modes of the HP phase of $SbPO_4$ have been measured and their frequencies have been compared to the theoretically predicted modes for the HP triclinic phase.

**Table 5.** Experimental and theoretical Raman mode frequencies and pressure coefficients of the triclinic HP phase ($P$-1) of $SbPO_4$ obtained by fitting the equation $\omega(P) = \omega_{10.9GPa}+a\cdot P$ from 10.9 GPa up to 14 GPa.

| Symmetry | Experimental | | Theoretical | |
|---|---|---|---|---|
| | $\omega_{10.9GPa}$ ($cm^{-1}$) | $a$ ($cm^{-1}$/GPa) | $\omega_{10.9GPa}$ ($cm^{-1}$) | $a$ ($cm^{-1}$/GPa) |
| $A_g$ | 106.7 | 1.0 | 95.0 | 0.7 |
| $A_g$ | 134.7 | 2.2 | 107.3 | 0.7 |
| $A_g$ | | | 110.8 | 1.4 |
| $A_g$ | | | 156.8 | 1.1 |
| $A_g$ | | | 166.8 | 0.6 |
| $A_g$ | | | 168.3 | 1.1 |
| $A_g$ | 223.3 | 3.8 | 216.5 | 2.1 |
| $A_g$ | 247.7 | 3.4 | 284.9 | 3.5 |
| $A_g$ | | | 326.6 | 1.5 |
| $A_g$ | 394.2 | 1.9 | 350.1 | 1.6 |
| $A_g$ | 452.0 | 2.4 | 492.1 | 2.4 |
| $A_g$ | 483.8 | 2.0 | 530.6 | -0.3 |
| $A_g$ | 555.2 | 2.0 | 538.6 | 1.0 |
| $A_g$ | 631.7 | 0.5 | 603.7 | 1.5 |
| $A_g$ | 1022.0 | 2.2 | 984.1 | 3.0 |
| $A_g$ | | | 996.1 | 3.9 |
| $A_g$ | | | 1004.3 | 2.4 |
| $A_g$ | 1118.6 | 4.0 | 1078.6 | 4.1 |

## 4.4. Electronic properties under compression

In order to understand the electronic properties of $SbPO_4$, we have calculated the theoretical electronic band structure of $SbPO_4$. **Figure 11** shows the theoretical electronic band structure and PDOS of $SbPO_4$ at 0 GPa and 5.1 GPa. As observed in **Figure 11(a)**, $SbPO_4$ presents a calculated indirect bandgap of 3.84 eV at 0 GPa, which valence band maximum (VBM) and conduction band minimum (CBM) are located at the $\mathbf{C_2}$ and **B** points of the BZ, respectively. A second minimum of the conduction band is located at the $\mathbf{Y_2}$ point of the BZ. Therefore, considering the underestimation of the bandgap from DFT-PBEsol calculations, the real bandgap must be well above 3.84 eV at room pressure. This means that $SbPO_4$ is an insulating and transparent material in the visible, UVA and UVB ranges. The maximum of the valence band is dominated by O states while the minimum of the conduction band is dominated by Sb states.

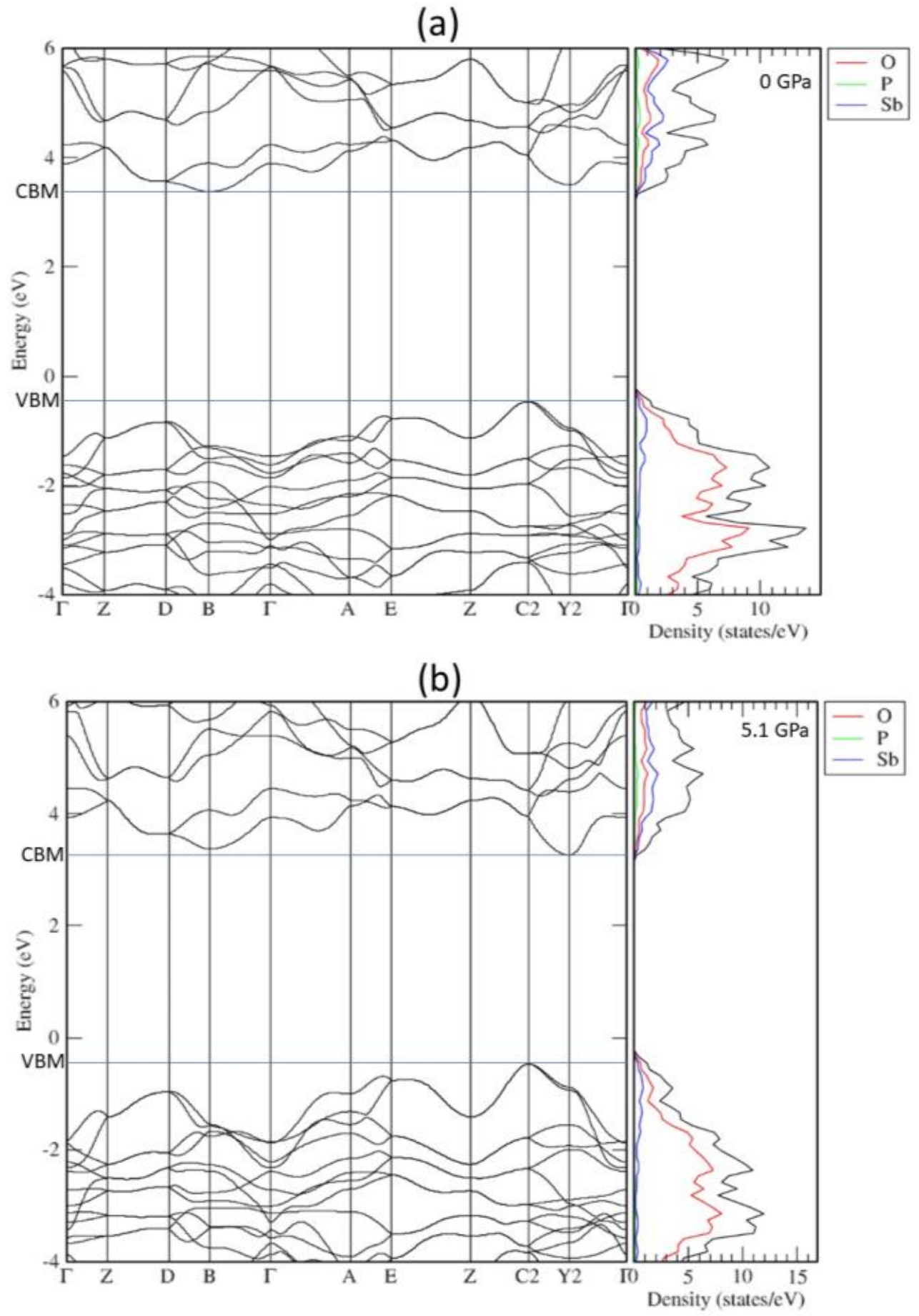

**Figure 11.** Theoretical electronic band structure of monoclinic $SbPO_4$ at (a) 0 GPa and (b) 5.1 GPa. The VBM and CBM lines indicate the Valence Band Maximum and Conduction Band Minimum, respectively.

In **Figure 11(b)** it is possible to observe that at 5.1 GPa, the minimum of the conduction band is located at point **$Y_2$**, indicating that the bandgap at this pressure range is measured between the high-symmetry points of **$C_2$** and **$Y_2$**. The pressure dependence of both indirect **$C_2$-B** and **$C_2$-$Y_2$** bandgaps is plotted in **Figure 12**. As can be observed, the indirect **$C_2$-B** bandgap increases with pressure whereas the indirect **$C_2$-$Y_2$** bandgap decreases with pressure. Consequently, an indirect-to-indirect crossover in the conduction band minimum occurs around 2.4 GPa; i.e. close to the IPT pressure. Above this pressure, the minimum indirect bandgap is found to be between the **$C_2$** and **$Y_2$** high-symmetry points of the BZ.

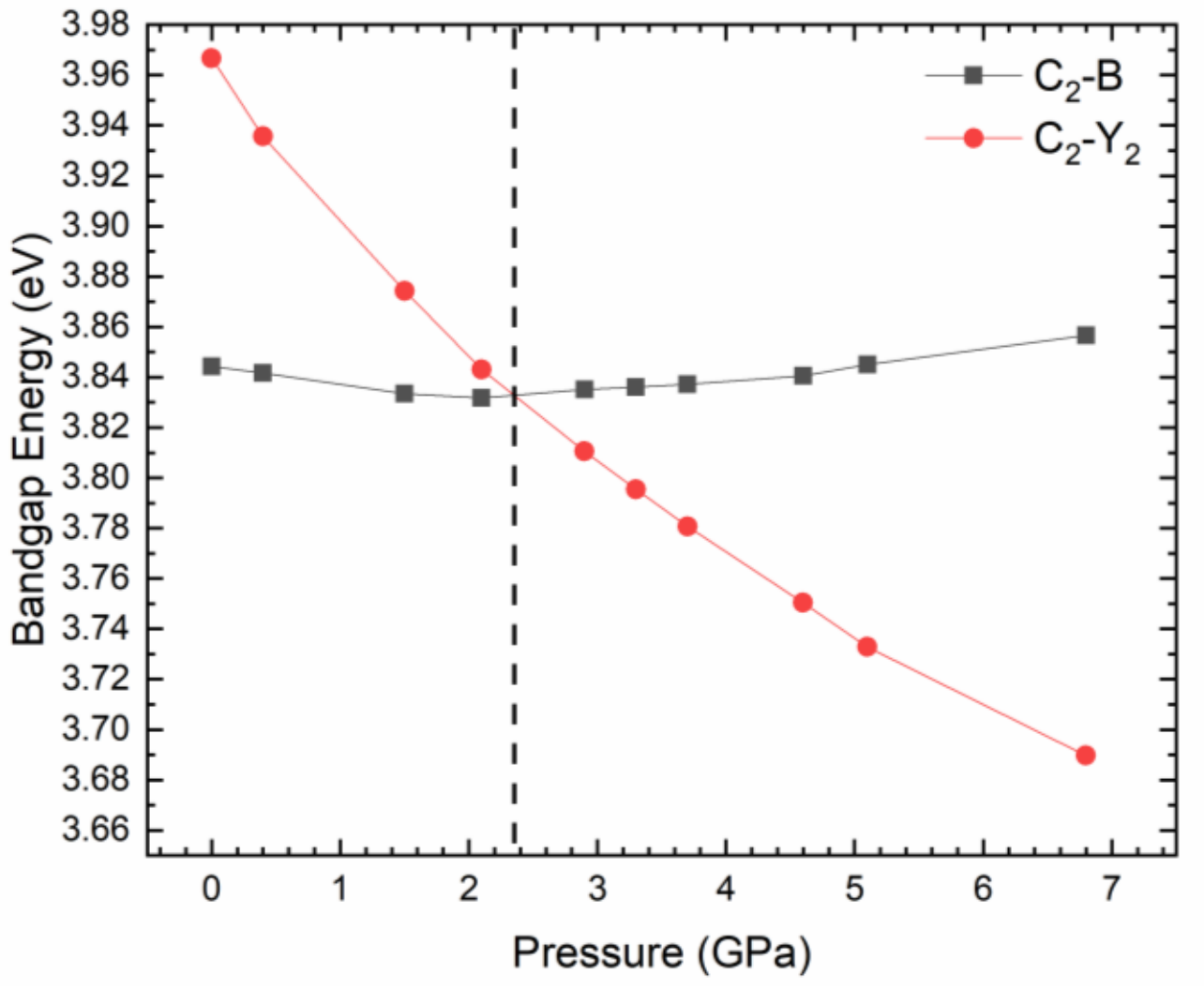

**Figure 12.** Pressure dependence of the theoretical indirect bandgaps **$C_2$-$Y_2$** and **$C_2$-B** in monoclinic $SbPO_4$ up to 6.8 GPa. The vertical dashed line indicates the pressure at which the crossing of the two indirect bandgaps occur.

To complete the picture of the evolution of $SbPO_4$ under compression and probe the variations of the Sb coordination as a function of pressure, we have performed an analysis of the ELF and the electron charge density using the QTAIM formalism for the different Sb-O bonds at different pressures in both monoclinic and triclinic $SbPO_4$ which can be compared to the ECoN (**Fig. 9(a)**). The ELF analysis is shown in **Fig. 13** where some remaining non-smoothness of the curves are due to the impossibility of raising the number of radial points further. We note that an all-electron wave function is needed to get a reliable picture of the ELF, since this function is not separable into core and valence contributions. The value of the ELF along lines connecting Sb to its O neighbors were calculated by three-dimensional interpolation from the ELF grid generated by Elk using the CRITIC2 software[49]. For the AIM electron density analysis, we have computed the electronic charge density and respective Laplacian at the BCPs also using CRITIC2 software (see **Table S5**). With this information, we have analyzed the Sb-O interatomic interactions in the different $SbPO_4$ structures in order to study the variation in Sb coordination as a function of pressure. At 0 GPa, Sb is 4-fold coordinated in monoclinic $SbPO_4$ with four Sb-O distances ($d_1$, $d_2$ and two $d_3$) below 2.2 Å (see **Fig. 6(a)**). All four bonds show a similar ELF profile and a minimum near 0.44 of the normalized distance in **Fig. 13(a)** and also similar values of the electron density at their BCPs. On the other hand, the remaining four Sb-O distances (two $d_4$, $d_5$ and $d_6$ above 2.7 Å) present completely different ELF profiles that show the existence of a maximum near 0.4 of the normalized distance that corresponds to the Sb LEP and a minimum close to 0.52 for $d_4$, 0.53 for $d_5$ and 0.54 for $d_6$. Regarding the electron density, the $d_4$ and $d_6$ Sb-O distances show charge densities at the corresponding BCPs that are significantly smaller than in the short contacts, and the $d_5$ contact does not even have a BCP (see **Table S5**). These observations evidence the negligible Sb-O interaction along these directions, which agrees with the ECoN results regarding the 4-fold coordination of Sb for the monoclinic $SbPO_4$ systems at 0 GPa.

Above 3 GPa, monoclinic $SbPO_4$ shows four Sb-O bond lengths ($d_1$ to $d_3$) below 2.2 Å and two Sb-O bonds ($d_4$) below 2.6 Å (see **Fig. 6(a)**). The four shortest distances show ELF profiles similar to those at 0 GPa and the other two distances ($d_4$) show an ELF profile where the LEP maximum is almost gone and there is a minimum closer to 0.44; i.e. similar to those of $d_1$ to $d_3$ distances (**Fig. 13(b)**); thus indicating that the ELF domain associated to the LEP has shrunk. Similarly, at this pressure, the charge density of the $d_4$ has increased significantly compared to the evolution of the density at the $d_1$, $d_2$, and $d_3$ BCPs. This picture of the ELF and charge density at the BCP is consistent with the 4+2-fold Sb coordination that occurs above the IPT. Moreover, the BCP along the $d_5$ distance appears at pressures above 3 GPa, thus giving support to the occurrence of an IPT above this pressure involving a change from 4-fold to 4+2-fold Sb coordination. Note that at 7.1 GPa the ELF of the $d_5$ and $d_6$ distances still show the LEP maxima near the 0.4 normalized distance (**Fig. 13 (b)**) and the charge densities at the BCPs of these two distances are smaller than the others (see **Table S5**), thus supporting the 4+2-fold coordination of $SbPO_4$ of the monoclinic phase up until 8 GPa. We also point out that the Laplacian of all BCPs is positive, thus evidencing the ionic character of all Sb-O bonds, regardless of bond distance.

Regarding the triclinic phase, we find four distances below 2.15 Å, two distances below 2.5 Å and the remaining two distances below 2.7 Å above 8 GPa. At 14.4 GPa, all $d_1$ to $d_8$ distances show ELF profiles (see **Figure 13(c)**) similar to those found in monoclinic $SbPO_4$ at 7.1 GPa (see **Figure 13(b)**). The degeneracy of bonds $d_3$ and $d_4$ in the monoclinic phase is broken in the triclinic phase, thus we find $d_1$ to $d_4$ ($d_5$ to $d_6$) distances in the triclinic phase showing similar ELF profiles to those of $d_1$ to $d_3$ ($d_4$) distances in the monoclinic phase. Similarly, $d_7$ and $d_8$ distances in the triclinic phase show similar ELF profiles to those of $d_5$ and $d_6$ in the monoclinic phase. This is consistent with the 4+2 coordination of Sb in triclinic $SbPO_4$ at 14.4 GPa.

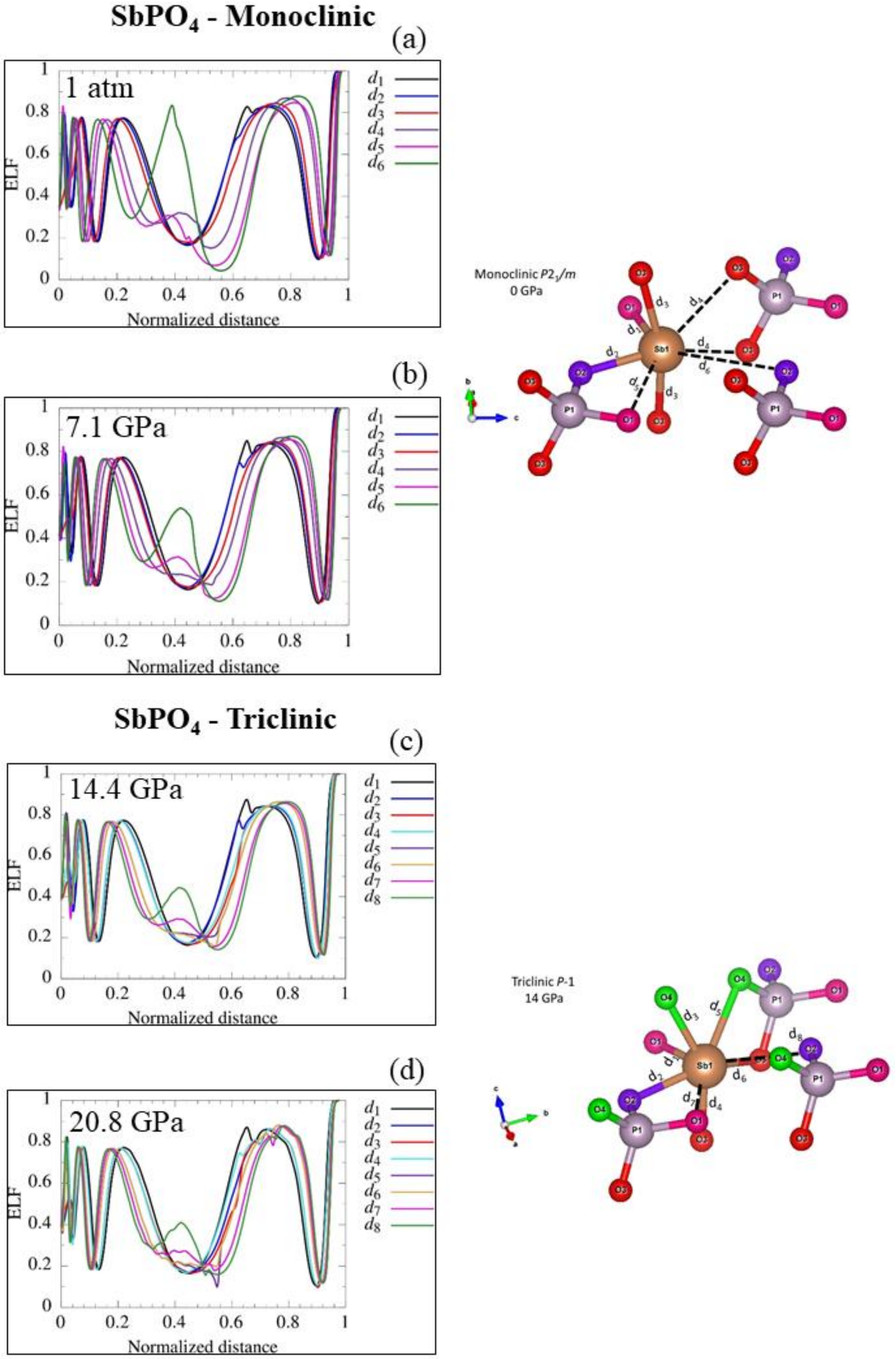

**Figure 13.** Theoretical all-electron 1D-ELF values along the eight shortest Sb-O distances (d1 to d6) in monoclinic $SbPO_4$ at 1 atm (a) and 7.1 GPa (b) and the same for the eight shortest Sb-O distances (d1 to d8) in triclinic $SbPO_4$ at 14.4 GPa (c) and 20.8 GPa (d). Note that both $d_3$ and $d_4$ distances are doubly degenerated in the monoclinic phase as shown in the structural detail.

At 20.8 GPa, the picture is slightly different because the ELF maximum due to the Sb LEP is gone for the $d_7$ distance; i.e. $d_7$ distance shows a similar ELF profile to that of $d_5$ and $d_6$ distances (see **Fig. 13 (d)**). This suggests an increase of coordination to 4+2+1 for Sb. This interpretation is in agreement with the fact that the charge density at the BCPs of the $d_7$ distance has values comparable to those in the $d_4$ distance of the monoclinic phase. Moreover, this conclusion is in agreement with the fact that the Sb coordination in $SbPO_4$ reaches the effective coordination found for Bi in $BiPO_4$-III at 0 GPa. Note that the charge density value of the $d_8$ distance is smaller than the others and the ELF profile of the $d_8$ distance still exhibits the maximum of the Sb LEP at 20.8 GPa. We interpret this as indicating that the triclinic structure Sb does not undergo a 4+2+1+1 coordination up to higher pressure (likely above ca. 25 GPa).

In summary, we have demonstrated with the calculated ELFs and the charge densities and respective Laplacians at the BCPs of the shortest Sb-O distances that in $SbPO_4$: i) a change in the number of BCPs of Sb occurs at the IPT close to 3 GPa, and ii) an increase of Sb coordination can be evidenced by the charge density accumulation at the BCPs, and by the disappearance of the Sb LEP maximum of the ELF, supporting the conclusion related to the increase of Sb coordination previously shown by the ECoN.

## 5. Conclusions

We have reported a joint experimental and theoretical study of the structural and vibrational properties of $SbPO_4$ at HP by means of XRD and RS measurements combined with *ab-initio* calculations. From the structural point of view, we have shown that $SbPO_4$ is one of the most compressible materials (bulk modulus around 20 GPa), not only among phosphates but also among $ABO_4$ compounds. Moreover, its compressibility tensor evidences a considerable anisotropic behavior due to a high non-linear compression, mainly along the *a*-axis. Additionally, our results have shown that $SbPO_4$ undergoes an IPT around 3 GPa and a PT above 9 GPa, which is completed around 20 GPa.

After the study of several candidates for the HP phase of $SbPO_4$ on the light of an updated Bastide's diagram containing many $ABO_4$ compounds with strong cation LEPs, we have proposed a triclinic distortion of the original monoclinic phase as the HP phase above 9 GPa. The Raman-active modes of both LP and HP phases have been measured and properly discussed at different pressures. In general, a rather good agreement is observed between the experimental and theoretical data for the structural and vibrational data. Finally, we have provided the electronic band structure of monoclinic $SbPO_4$ at different pressures, showing that this compound is an indirect bandgap material (bandgap value above 3.8 eV) that is transparent in the visible, UVA and UVB spectral regions in the whole pressure range up to 9 GPa.

Theoretical data have helped us to understand the microscopic mechanisms of the compression of monoclinic $SbPO_4$, evidencing that monoclinic $SbPO_4$ undergoes a transition from 2D-type structure with a 4-fold coordination of Sb at room pressure to a 3D structure with 4+2 coordination above 3 GPa. Changes of the Wyckoff positions, changes of the slopes of c/a and b/a ratios, and changes on the pressure dependence of interatomic distances, even of P-O3 bond distances (expected to be rather strong and incompressible bonds), clearly show the occurrence of an IPT around 3 GPa. This IPT is further confirmed by the changes in the pressure coefficients of different vibrational modes around 3 GPa. Moreover, all vibrational modes of $SbPO_4$ that show a negative pressure coefficient at room pressure change to a positive pressure coefficient above 3 GPa. This result is in good agreement with the pressure-induced 2D-to-3D phase transition taking place in layered $SbPO_4$ above 3 GPa. The Sb cation increases the coordination number up to 4+2+1-fold for the triclinic phase above 15 GPa and the effective coordination of $SbPO_4$ around 18 GPa becomes similar to $BiPO_4$-III at room pressure.

Finally, we want to stress that the ability of pressure to modulate the LEP activity and convert the 2D structure of $SbPO_4$ into the 3D network of $BiPO_4$-III may have important implications for technological applications for $SbPO_4$-based compounds since the role played by external pressure can be mimicked by chemical pressure. In particular, partial substitution of Sb cations in $SbPO_4$ by Bi cations (with smaller LEP) or by other cations with valence 3+ and without an active LEP, like In, is expected to lead to a closing of the inter-layer space of the $SbPO_4$ structure; i.e. will promote the 3D nature of the compound. Conversely, partial substitution of Sb cations by As cations (with a much stronger LEP) is expected to promote the opening of the structure and consequently the 2D nature of the compound. Thus, our work suggests a way to open or close the structure of layered $SbPO_4$ that can help to enhance the catalytic and atomic-insertion properties of $SbPO_4$-based compounds

## 6. Acknowledgments

Authors thank the financial support from Brazilian Conselho Nacional de Desenvolvimento Científico e Tecnológico (CNPq - 159754/2018-6, 307199/2018-5, 422250/2016-3, 201050/2012-9), FAPESP (2013/07793-6), Spanish Ministerio de Economia y Competitividad (MINECO) under projects MALTA Consolider Ingenio 2010 network (MAT2015-71070-REDC and RED2018-102612-T), MAT2016-75586-C4-1/2/3-P, PGC2018-097520-A-100, FIS2017-83295-P, and PGC2018-094417-B-I00 from Generalitat Valenciana under project PROMETEO/2018/123 and from European Comission under project COMEX. D. S.-P., J. A. S., and A.O.R. acknowledge “Ramón y Cajal” Fellowship for financial support (RyC-2014-15643, RYC-2015-17482, and RyC-2016-20301 respectively). E. L. d.-S., A. M., A. B. and

P. R.-H. acknowledge computing time provided by Red Española de Supercomputación (RES) and MALTA-Cluster.

**Supporting Information Available:** Theoretical atomic coordinates of monoclinic ($P2_1/m$) and high-pressure triclinic ($P$-1) $SbPO_4$; Representation of the $SbPO_4$ atomic vibration; Angle-dispersive XRD and Raman spectra of $SbPO_4$ measured at different pressures; Detail of the monoclinic structure of $SbPO_4$ along the ac-plane; Evolution of the theoretical Wyckoff positions of monoclinic $SbPO_4$ with pressure; Theoretical enthalpy difference vs pressure; Evolution of the Sb eccentricity in the SbO6 polyhedron of monoclinic $SbPO_4$; Theoretical pressure dependence of the IR-active modes of monoclinic and triclinic $SbPO_4$; Table with Sb-O distances, charge density, and its Laplacian at the BCPs of the different Sb-O distances in the monoclinic and triclinic phase of $SbPO_4$ at different pressures.

## For Table of Contents Only

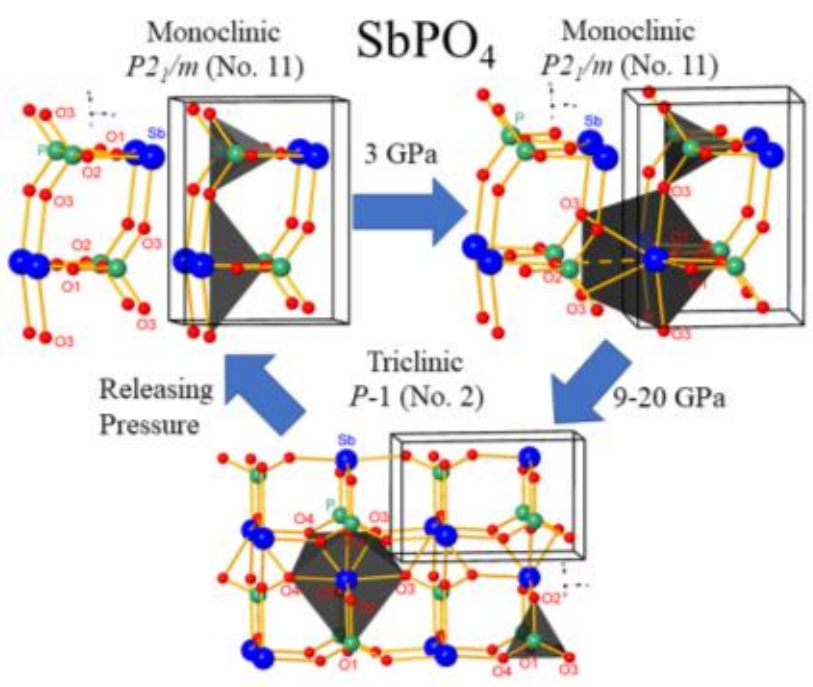

**TOC figure**

### Synopsis

Here we report a joint experimental and theoretical study of the monoclinic $SbPO_4$ at high pressure. We show that $SbPO_4$ is not only one of the most compressible phosphates but also one of the most compressible compounds of the $ABO_4$ family. The strong compression along *a*-axis leads to an isostructural phase transition above 3 GPa and a reversible phase transition to a triclinic phase above 9 GPa, which is completed above 20 GPa.

**Supplementary Material of**

# Experimental and theoretical study of $SbPO_4$ under compression

**André Luis de Jesus Pereira,[1,2*] David Santamaría-Pérez,[3] Rosário Vilaplana,[4] Daniel Errandonea,[3] Catalin Popescu,[5] Estelina Lora da Silva,[1] Juan Angel Sans,[1] Juan Rodríguez-Carvajal,[6] Alfonso Muñoz,[7] Plácida Rodríguez-Hernández,[7] Andres Mujica,[7] Silvana Elena Radescu,[7] Armando Beltrán,[8] Alberto Otero de la Roza,[9] Marcelo Nalin,[10] Miguel Mollar,[1] and Francisco Javier Manjón[1*]**

*[1] Instituto de Diseño para la Fabricación y Producción Automatizada, MALTA Consolider Team, Universitat Politècnica de València, València, Spain*

*[2] Grupo de Pesquisa de Materiais Fotonicos e Energia Renovável - MaFER, Universidade Federal da Grande Dourados, Dourados, MS, Brazil*

*[3] Departament de Física Aplicada – ICMUV, MALTA Consolider Team, Universitat de València, Burjassot, Spain*

*[4] Centro de Tecnologías Físicas, MALTA Consolider Team, Universitat Politecnica de València, València 46022, Spain*

*[5] CELLS-ALBA Synchrotron Light Facility, 08290 Cerdanyola, Barcelona, Spain*

*[6] Institut Laue-Langevin, 71 Avenue des Martyrs, CS 20156, 38042, Grenoble, Cedex 9, France*

*[7] Departamento de Física, Instituto de Materiales y Nanotecnología, MALTA Consolider Team, Universidad de La Laguna, Tenerife, Spain*

*[8] Departament de Química Física, MALTA Consolider Team, Universitat de València, 46100 Burjassot, Spain*

*[9] Departamento de Química Física y Analítica, MALTA Consolider Team, Universidad de Oviedo, 33006 Oviedo, Spain*

*[10] Instituto de Quimica, Depto. Química Geral e Inorgânica, UNESP - Campus de Araraquara, SP, Brazil*

## Structural and Vibrational Properties at Room Pressure

**Table S1.** Theoretical atomic coordinates of the monoclinic $P2_1/m$ (space group Nr. 11) structure of $SbPO_4$ at room conditions. Lattice parameters are: $a$ = 5.0482 Å, $b$ = 6.8228 Å, $c$ = 4.7392 Å and $\beta$ = 97.0624°, with a unit cell volume $V_0$= 161.9968 Å$^3$.

| Atom | Wyckoff position | *x* | *y* | *z* |
|---|---|---|---|---|
| Sb | 2e | 0.16677 | 0.25 | 0.18534 |
| P | 2e | 0.61036 | 0.25 | 0.72021 |
| O1 | 2e | 0.32278 | 0.25 | 0.81191 |
| O2 | 2e | 0.55847 | 0.25 | 0.38998 |
| O3 | 4f | 0.77615 | 0.07144 | 0.83277 |

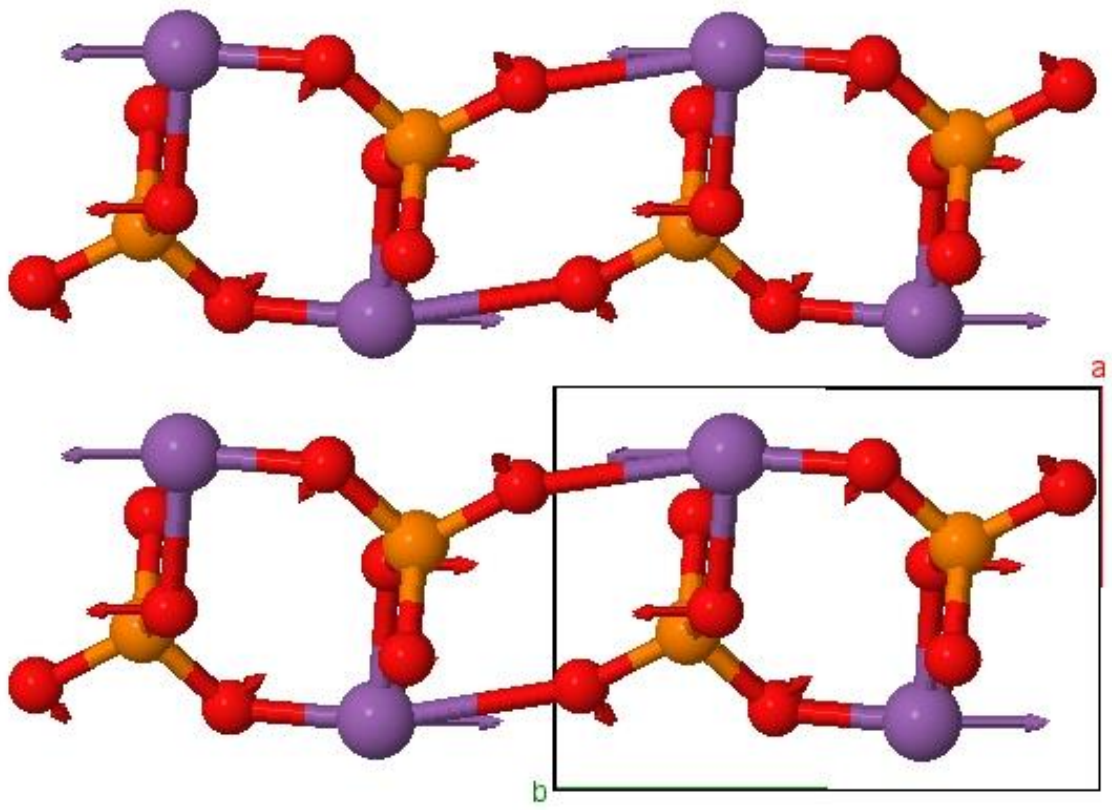

**Figure S1.** Atomic vibrations (along the *b*-axis) of the $B_g$ mode of 75 $cm^{-1}$, which is one of the shear layer modes of $SbPO_4$. Sb (big purple), P (medium-size orange) and O (small red).

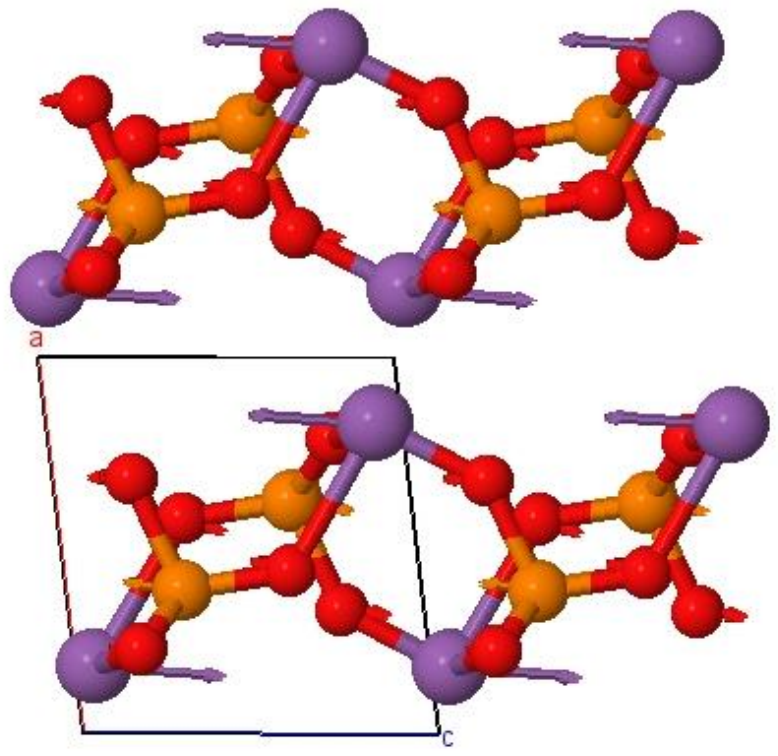

**Figure S2.** Atomic vibrations (along the *c*-axis) of the $A_g$ mode of 89 $cm^{-1}$, which is one of the shear layer modes of $SbPO_4$. Sb (big purple), P (medium-size orange) and O (small red).

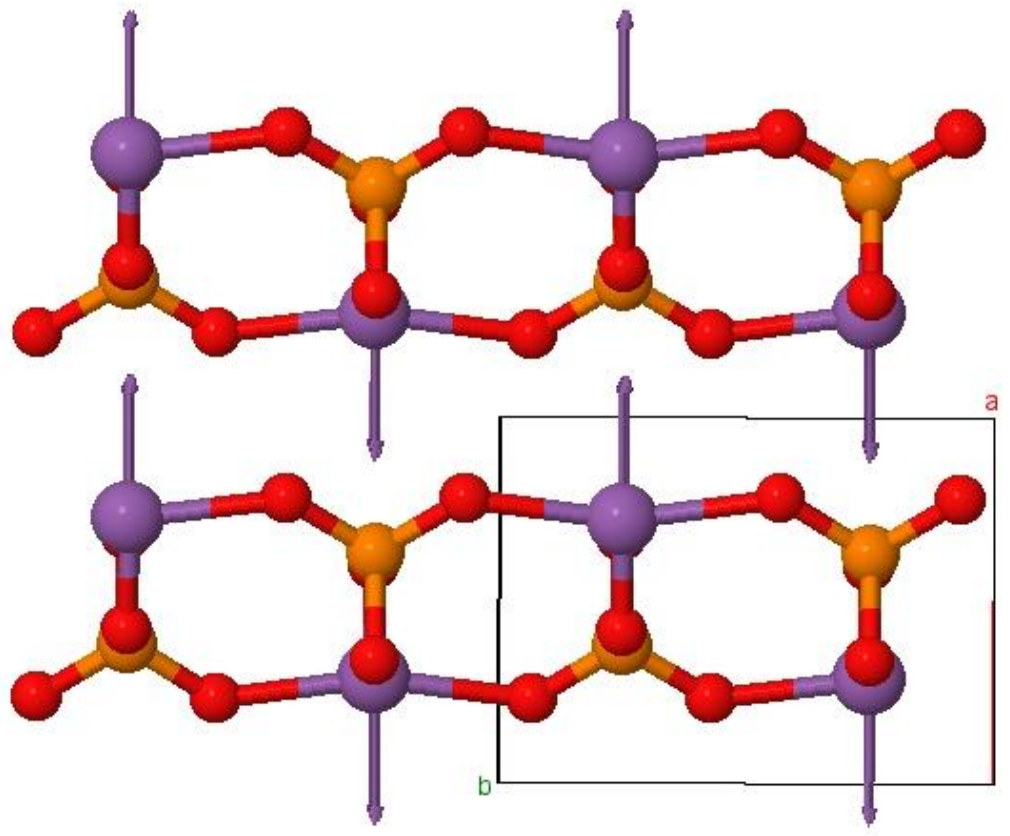

**Figure S3.** Atomic vibrations (along the *a*-axis) of the $A_g$ mode of 106 $cm^{-1}$, which is the main longitudinal layer mode of $SbPO_4$. Sb (big purple), P (medium-size orange) and O (small red).

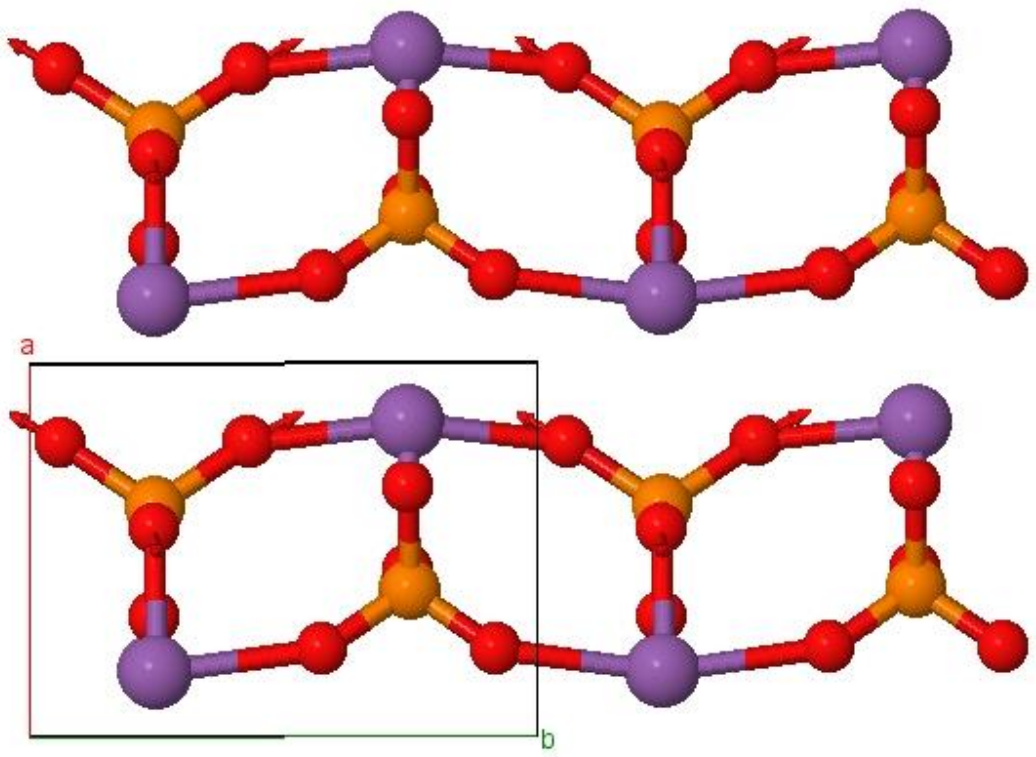

**Figure S4.** Atomic vibrations (mainly of O atoms) of the $A_g$ mode of 936 $cm^{-1}$, which is the symmetric stretching mode of $PO_4$ in $SbPO_4$. Sb (big purple), P (medium-size orange) and O (small red).

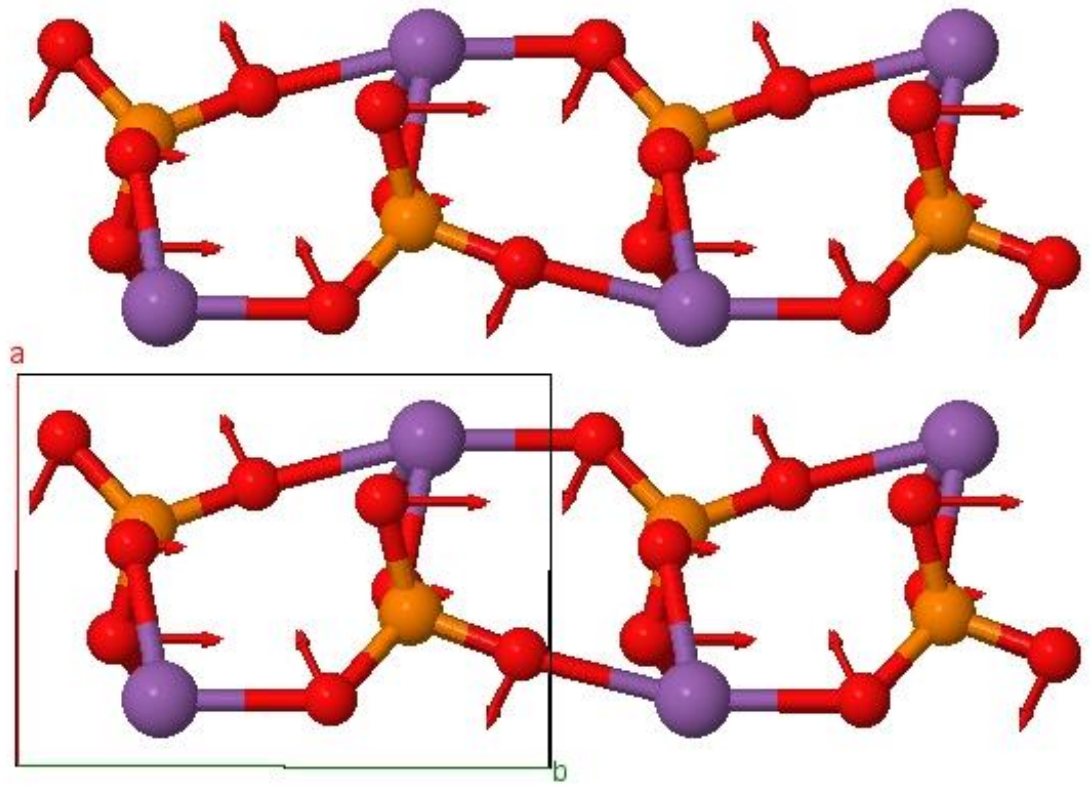

**Figure S5.** Atomic vibrations of the $A_u$ mode of 220 $cm^{-1}$, which is the main rotational mode of $PO_4$ in $SbPO_4$. Sb (big purple), P (medium-size orange) and O (small red).

## Structural properties of $SbPO_4$ at high pressure

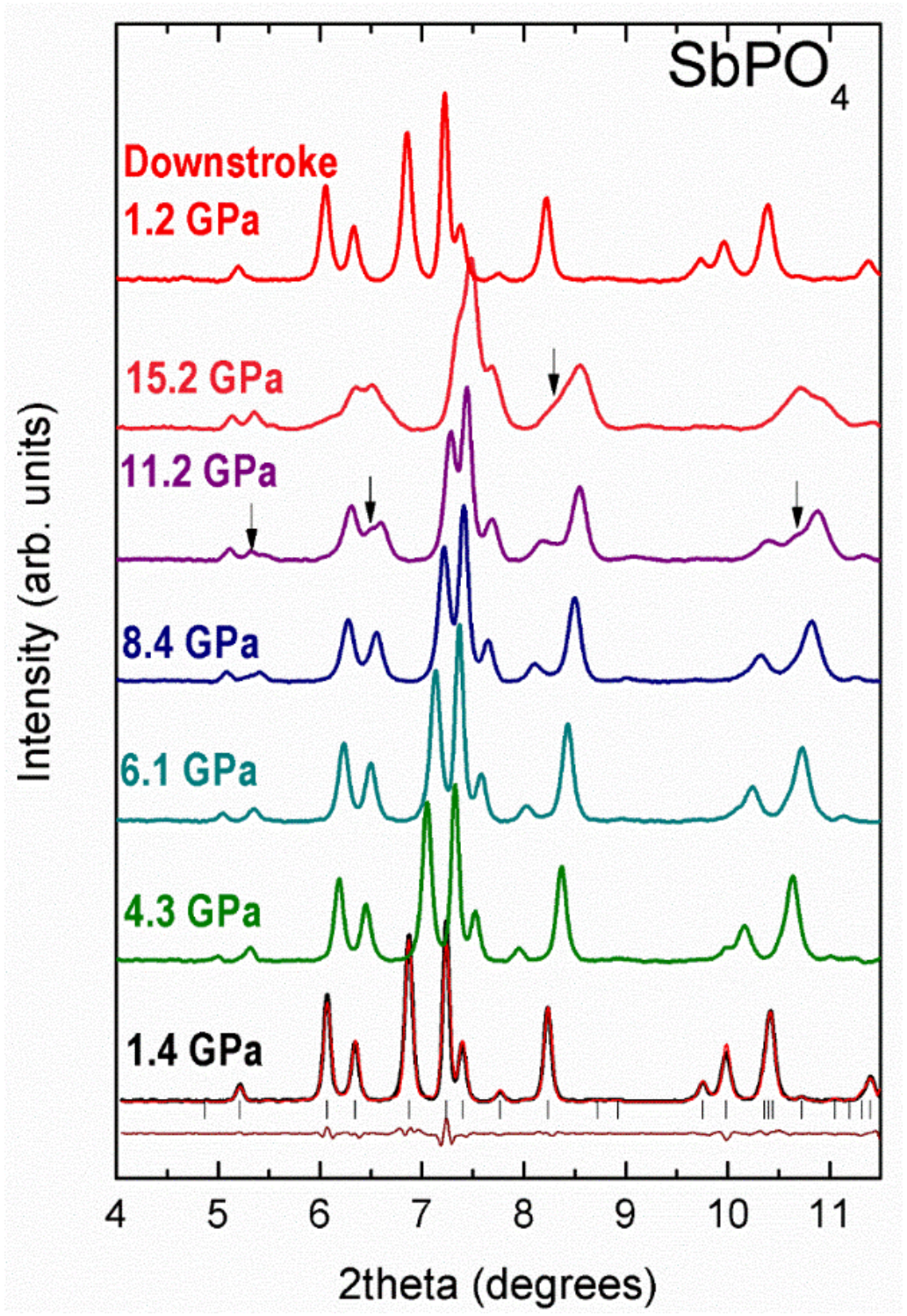

**Figure S6.** Angle-dispersive XRD of $SbPO_4$ measured at different pressures up to 15.2 GPa at room temperature. In order to facilitate the identification of changes in the XRD pattern related to the phase transition we represent $2\theta < 11.5^o$; i.e. the part of the XRD without overlapping with Cu or gasket peaks. The top pattern corresponds to the recovered sample at 1.2 GPa after decompression from 15.2 GPa. The black arrows indicate the position of new peaks not related to the initial phase.

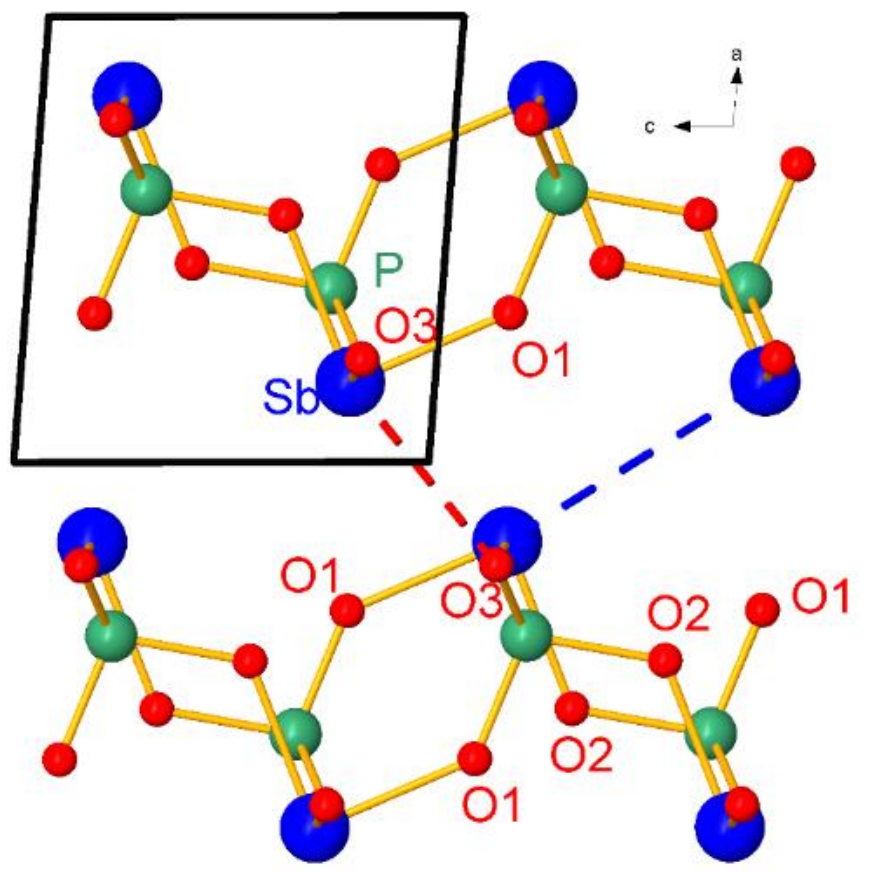

**Figure S7.** Detail of the monoclinic structure of $SbPO_4$ along the *ac*-plane. The directions of maximum and intermediate compressibility at room pressure are approximately along the plotted dashed red and blue lines, respectively.

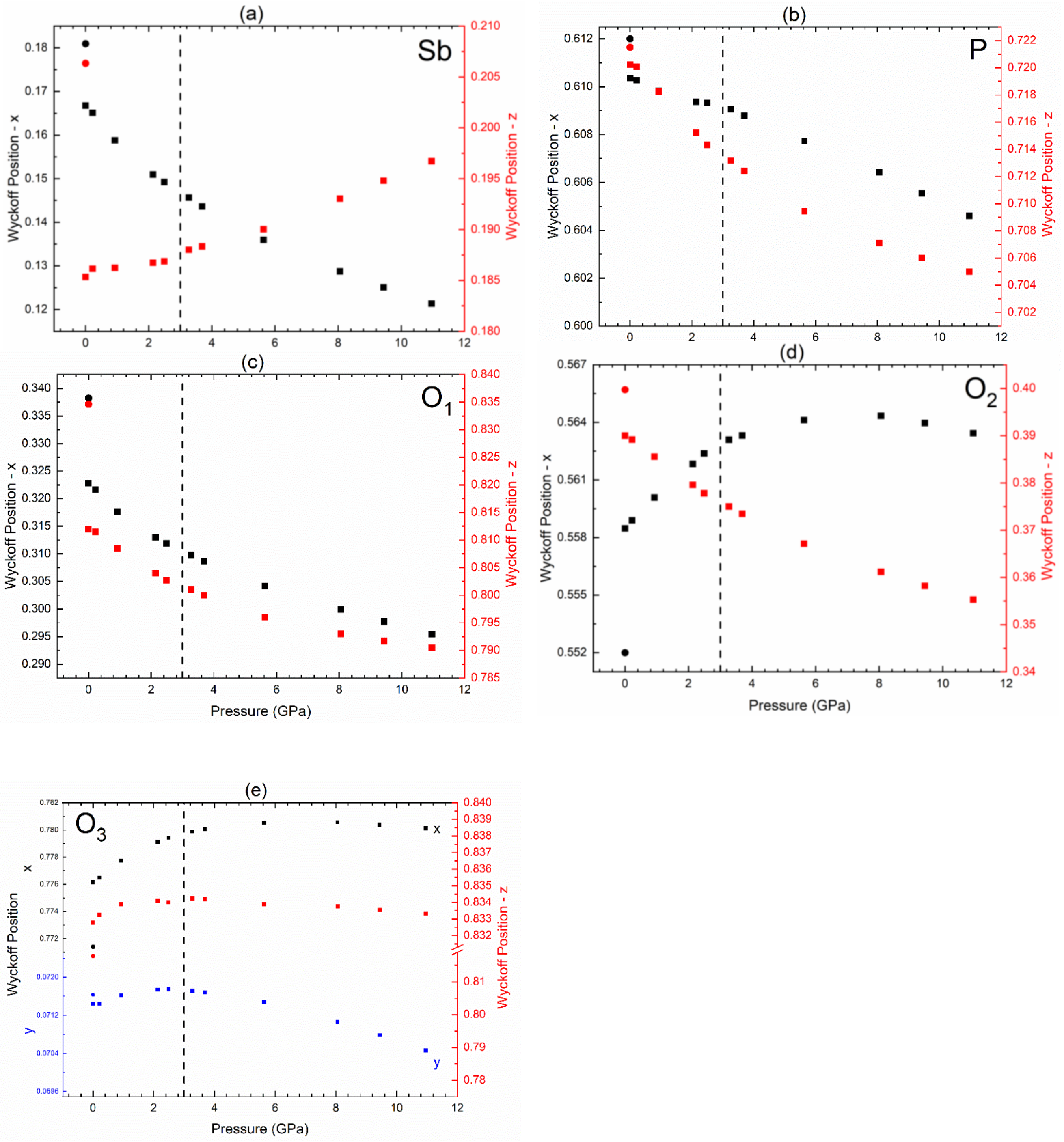

**Figure S8.** Evolution of the theoretical Wyckoff positions of monoclinic $SbPO_4$ with pressure (squares). The circles represent the experimental positions obtained by the Rietveld refinement of the XRD measurement performed at room pressure. The vertical dashed lines at 3 GPa indicate the pressure at which the IPT occurs as suggested by the change of the tendency of many Wyckoff coordinates.

**Table S2.** Theoretical (GGA-PBESol) atomic coordinates of proposed triclinic *P-1* (space group Nr. 2) structure of $SbPO_4$ at 15.005 GPa. Lattice parameters are: *a* = 4.3335 Å, *b* = 4.7040 Å, *c* = 6.5659 Å, $\alpha$ = 89.8023º, $\beta$ = 89.6317º, and $\gamma$ = 85.3338º, with a unit cell volume $V_0$= 133.3999 Å$^3$.

| Atom | Wyckoff position | *x* | *y* | *z* |
|---|---|---|---|---|
| Sb | 2i | 0.79835 | 0.11293 | 0.75058 |
| P | 2i | 0.29724 | 0.60217 | 0.74951 |
| O1 | 2i | 0.21180 | 0.29039 | 0.74841 |
| O2 | 2i | 0.65143 | 0.56153 | 0.74953 |
| O3 | 2i | 0.16606 | 0.77782 | -0.06971 |
| O4 | 2i | 0.83102 | 0.21922 | 0.43054 |

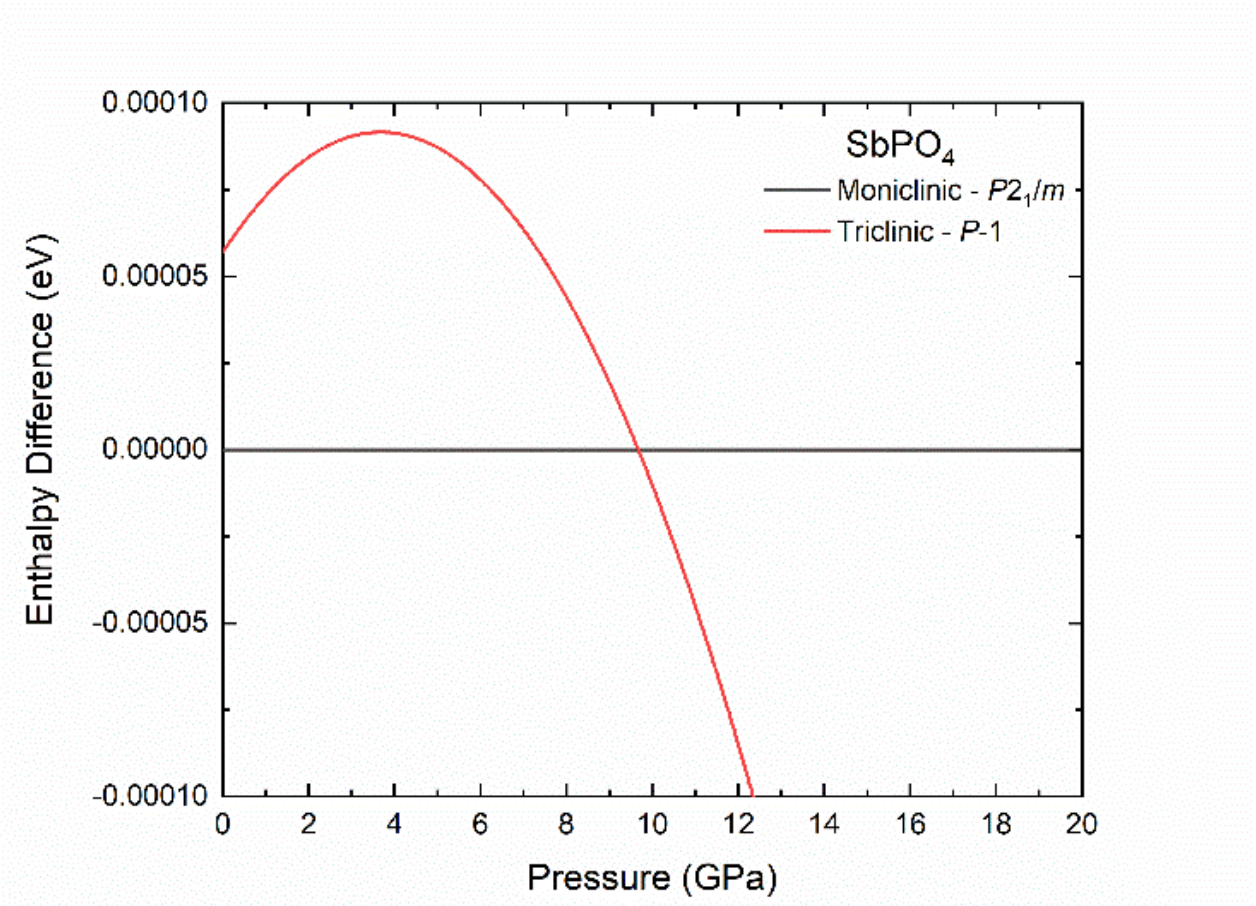

**Figure S9.** Theoretical enthalpy difference vs pressure of the triclinic HP phase of $SbPO_4$ (red line) with respect to the monoclinic LP phase (black line).

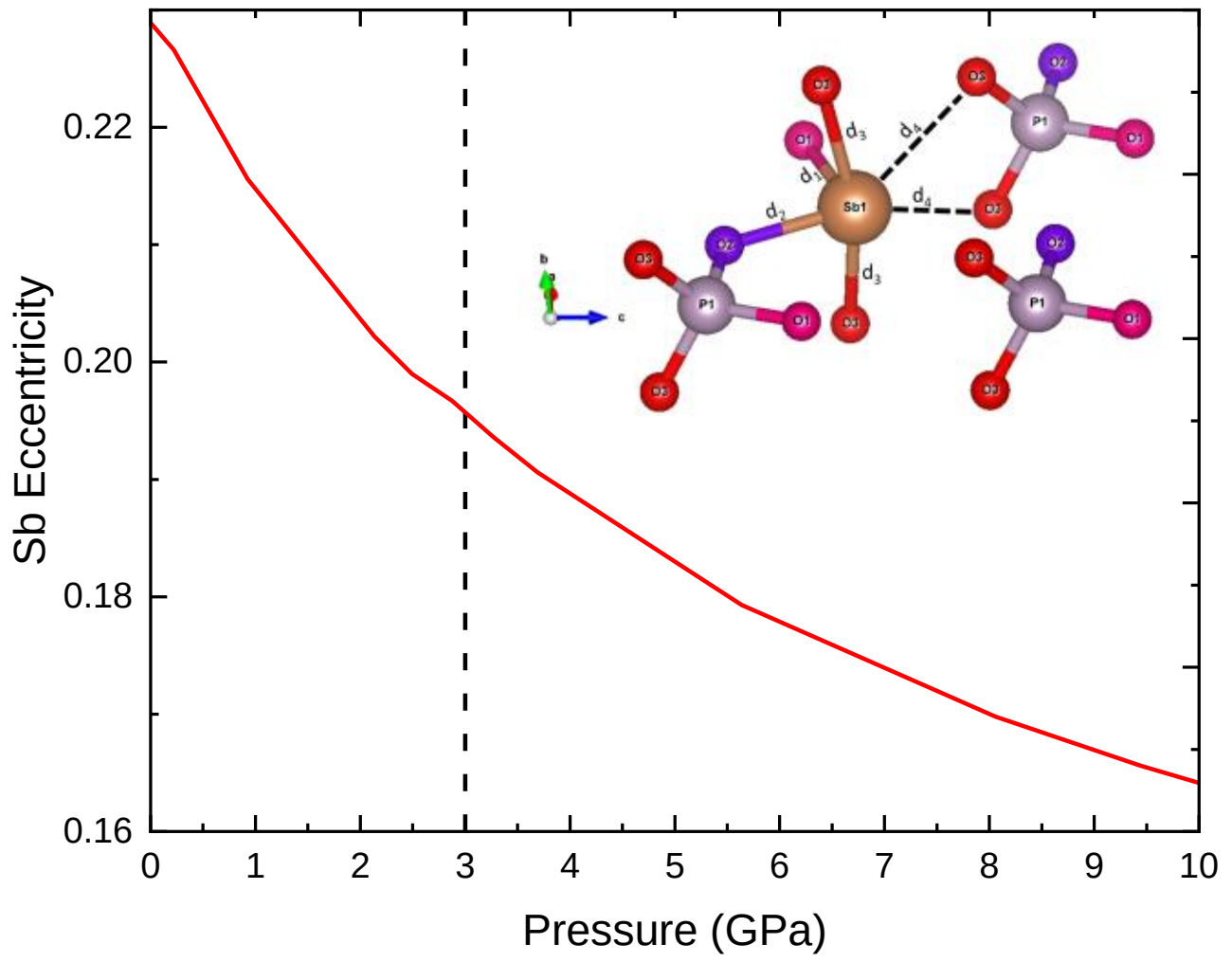

**Figure S10.** Pressure dependence of the theoretical Sb eccentricity in the SbO6 polyhedron of monoclinic $SbPO_4$. The Sb eccentricity was calculated using the IVTON software[1].

## Vibrational properties of $SbPO_4$ at high pressure

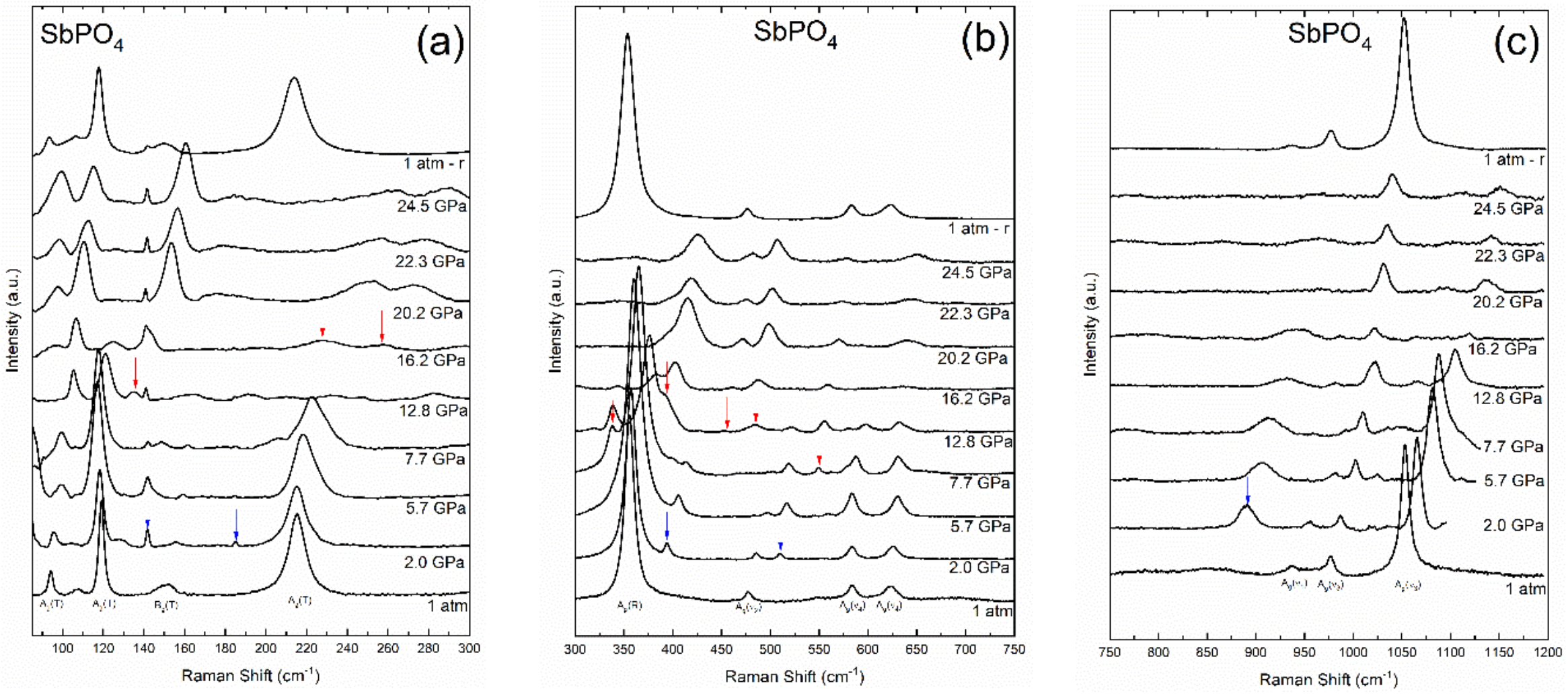

**Figure S11.** Room-temperature Raman spectra of $SbPO_4$ at selected pressures up to 24.5 GPa. (a) Low-frequency region, (b) Middle-frequency region, and (c) High-frequency region. Top Raman spectrum corresponds to the recovered sample after decompression from 24.5 GPa. Red arrows indicate the position of new peaks not related to the initial phase and blue arrows indicate peaks that either are not related to the sample or correspond to second-order modes of the original sample (see main text). In the bottom of the figures we have added a tentative mode assignment of the initial sample based in the theoretical results and the pressure evolution of these vibrational modes. Notations T, R, $\nu_1$, $\nu_2$, $\nu_3$, and $\nu_4$ refer to the main character of translation, rotation, or internal modes of the $PO_4$ units, respectively.

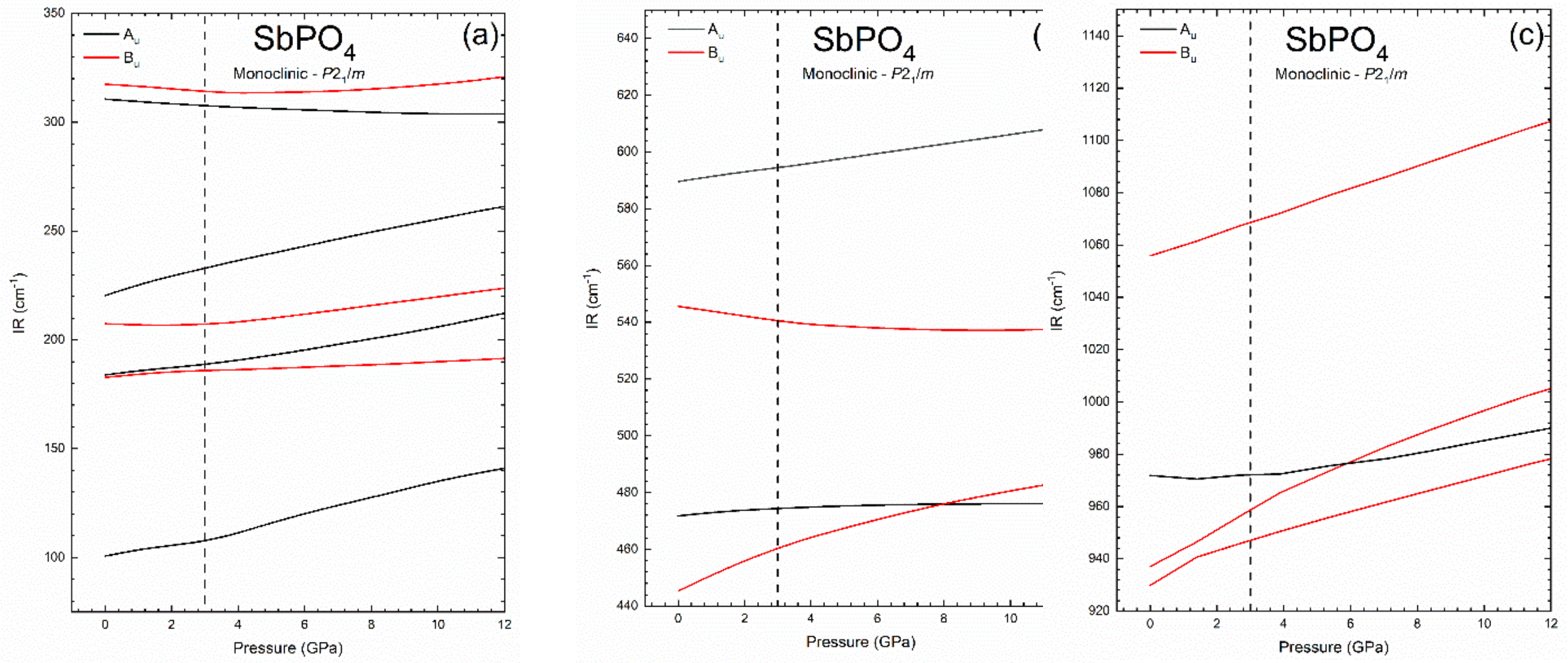

**Figure S12.** Theoretical pressure dependence of the IR-active modes of monoclinic $SbPO_4$: (a) from 75 to 350 $cm^{-1}$; (b) from 440 to 620 $cm^{-1}$, (c) from 920 to 1150 $cm^{-1}$. The vertical dashed lines at 3 GPa indicate the pressure at which the IPT occurs as suggested by the change of many frequency pressure coefficients.

**Table S3.** Theoretical IR-active mode frequencies and pressure coefficients of monoclinic $SbPO_4$ obtained by fitting the equation $\omega(P) = \omega_0 + a \cdot P$ up to 3 GPa.

| | Theoretical | |
|---|---|---|
| Symmetry | $\omega_0$ ($cm^{-1}$) | $a$ ($cm^{-1}$/GPa) |
| $A_u$ (T) | 101(2) | 2.2(4) |
| $B_u$ (T) | 183(2) | 1.2(2) |
| $A_u$ (R) | 184(2) | 1.6(2) |
| $B_u$ (T) | 207(3) | -0.2(2) |
| $A_u$ (R) | 220(3) | 4.2(4) |
| $A_u$ ($\nu_2$) | 311(4) | -1.1(1) |
| $B_u$ (R) | 317(4) | -1.1(1) |
| $B_u$ ($\nu_2$) | 445(5) | 5.2(2) |
| $A_u$ ($\nu_4$) | 472(5) | 0.9(1) |
| $B_u$ ($\nu_4$) | 546(6) | -1.7(1) |
| $B_u$ ($\nu_4$) | 589(6) | 1.7(1) |
| $B_u$ ($\nu_3$) | 930(8) | 6.0(1.0) |
| $B_u$ ($\nu_1$) | 937(8) | 7.2(3) |
| $A_u$ ($\nu_3$) | 972(8) | 0.04(0.60) |
| $B_u$ ($\nu_3$) | 1056(9) | 4.3(2) |

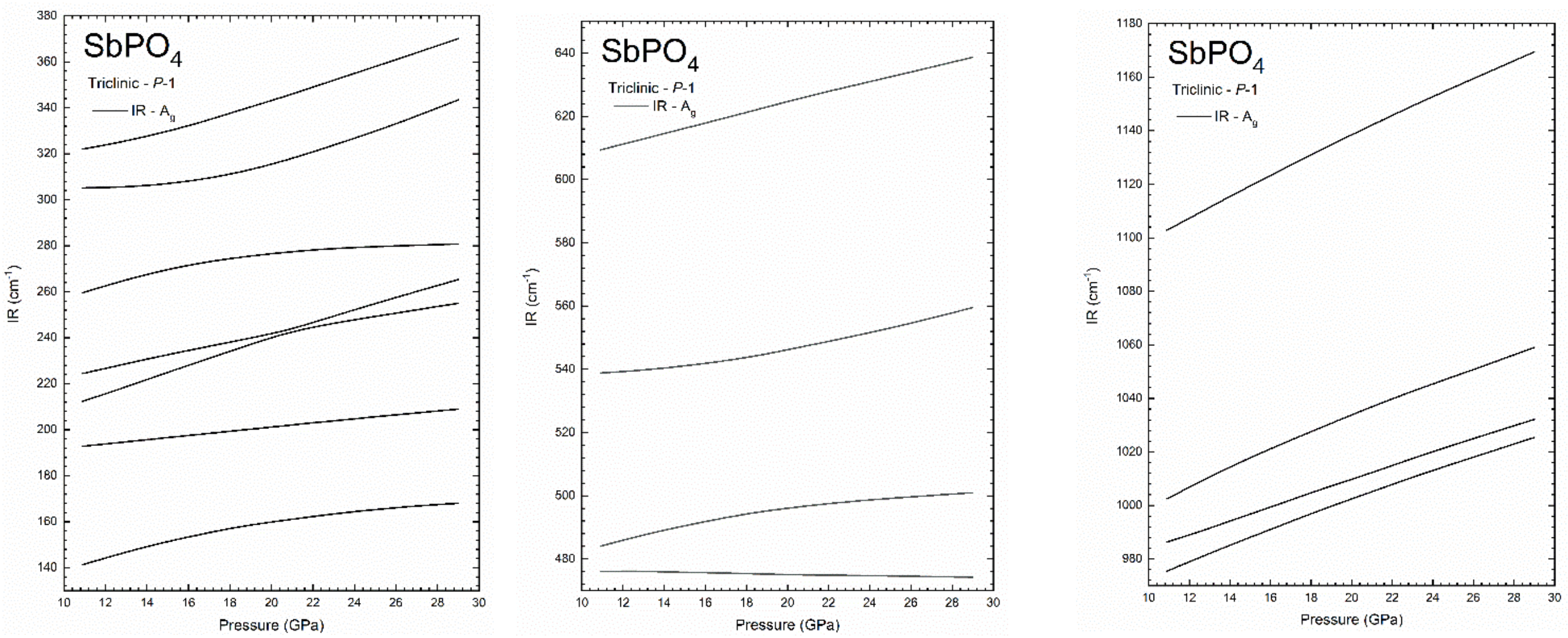

**Figure S13.** Pressure dependence of the theoretical IR-active modes of the triclinic HP ($P$-1) phase of $SbPO_4$: (a) from 130 to 380 $cm^{-1}$; (b) from 470 to 650 $cm^{-1}$, (c) from 970 to 1180 $cm^{-1}$.

**Table S4.** Theoretical IR-active mode frequencies and pressure coefficients of the triclinic HP (*P*-1) phase of $SbPO_4$ obtained by fitting the equation $\omega(P) = \omega_{10.9GPa} + a \cdot P$ from 10.9 GPa up to 14 GPa.

| Symmetry | Theoretical $\omega_{10.9GPa}$ ($cm^{-1}$) | Theoretical $a$ ($cm^{-1}$/GPa) |
|---|---|---|
| $A_u$ | 141.3 | 2.53 |
| $A_u$ | 192.8 | 0.91 |
| $A_u$ | 212.3 | 3.07 |
| $A_u$ | 224.4 | 2.03 |
| $A_u$ | 259.6 | 2.56 |
| $A_u$ | 305.2 | 0.36 |
| $A_u$ | 322.1 | 1.83 |
| $A_u$ | 476.0 | -0.04 |
| $A_u$ | 484.0 | 1.63 |
| $A_u$ | 538.8 | 0.52 |
| $A_u$ | 609.4 | 1.66 |
| $A_u$ | 975.5 | 3.11 |
| $A_u$ | 986.4 | 2.55 |
| $A_u$ | 1002.4 | 3.82 |
| $A_u$ | 1102.9 | 4.06 |

**Table S5.** Sb-O distances, charge density, and its Laplacian at the BCPs of the different Sb-O distances calculated at different pressures for the monoclinic and triclinic phases of $SbPO_4$.

| distance | l(Sb-O) [Å] | $\rho(\mathbf{r})$ [A.U.] | $\nabla^2\rho(\mathbf{r})$ [A.U.] |
|---|---|---|---|
| | Monoclinic 0 GPa | | |
| $d_1$ | 2.02339 | 0.11079 | 0.37097 |
| $d_2$ | 2.09001 | 0.09763 | 0.27354 |
| $d_3$ (x2) | 2.21486 | 0.07231 | 0.16590 |
| $d_4$ (x2) | 2.72456 | 0.02811 | 0.06543 |
| $d_6$ | 3.35994 | 0.00915 | 0.02447 |
| | Monoclinic 1.9 GPa | | |
| $d_1$ | 2.02878 | 0.10964 | 0. 30392 |
| $d_2$ | 2.11659 | 0.09261 | 0. 23692 |
| $d_3$ (x2) | 2.21421 | 0.07231 | 0.16348 |
| $d_4$ (x2) | 2.61149 | 0.03483 | 0.07713 |
| $d_6$ | 3.12937 | 0.01403 | 0.03199 |
| | Monoclinic 2.6 GPa | | |
| $d_1$ | 2.03018 | 0.10933 | 0.28307 |
| $d_2$ | 2.12284 | 0.09146 | 0.25143 |
| $d_3$ (x2) | 2.21295 | 0.07246 | 0.16420 |
| $d_4$ (x2) | 2.58776 | 0.03645 | 0.08129 |
| $d_6$ | 3.08092 | 0.01534 | 0.03579 |
| | Monoclinic 3.5 GPa | | |
| $d_1$ | 2.03136 | 0.10906 | 0.28190 |
| $d_2$ | 2.12786 | 0.09056 | 0.16683 |
| $d_3$ (x2) | 2.21006 | 0.07286 | 0.16658 |
| $d_4$ (x2) | 2.56892 | 0.03777 | 0.08368 |
| $d_5$ | 2.82774 | 0.02322 | 0.07290 |
| $d_6$ | 3.03545 | 0.01669 | 0.04056 |
| | Monoclinic 4.4 GPa | | |
| $d_1$ | 2.03257 | 0.10877 | 0.27837 |
| $d_2$ | 2.13330 | 0.08957 | 0.18874 |
| $d_3$ (x2) | 2.20692 | 0.07330 | 0.16912 |
| $d_4$ (x2) | 2.54936 | 0.03921 | 0.08624 |
| $d_5$ | 2.80201 | 0.02434 | 0.07734 |
| $d_6$ | 2.99071 | 0.01813 | 0.04496 |
| | Monoclinic 5.4 GPa | | |
| $d_1$ | 2.03317 | 0.10861 | 0.27036 |
| $d_2$ | 2.13794 | 0.08871 | 0.21201 |
| $d_3$ (x2) | 2.20384 | 0.07372 | 0.17080 |
| $d_4$ (x2) | 2.53247 | 0.04050 | 0.08879 |
| $d_5$ | 2.77898 | 0.02544 | 0.07566 |
| $d_6$ | 2.94995 | 0.01956 | 0.05080 |
| | Triclinic 9.2 GPa | | |
| $d_1$ | 2.03632 | 0.10780 | 0.27534 |
| $d_2$ | 2.14858 | 0.08681 | 0.23149 |
| $d_3$ | 2.18742 | 0.07614 | 0.18776 |
| $d_4$ | 2.19111 | 0.07553 | 0.17683 |
| $d_5$ | 2.48779 | 0.04410 | 0.09570 |
| $d_6$ | 2.49095 | 0.04389 | 0.10101 |
| $d_7$ | 2.69777 | 0.02975 | 0.08474 |
| $d_8$ | 2.83082 | 0.02444 | 0.06802 |
| | Triclinic 14.8 GPa | | |
| $d_1$ | 2.03757 | 0.10734 | 0.30906 |
| $d_2$ | 2.15416 | 0.08580 | 0.21052 |
| $d_3$ | 2.16514 | 0.07958 | 0.21339 |
| $d_4$ | 2.16902 | 0.07892 | 0.19852 |
| $d_5$ | 2.45254 | 0.04716 | 0.10237 |
| $d_6$ | 2.45335 | 0.04716 | 0.10157 |
| $d_7$ | 2.61003 | 0.03529 | 0.08232 |
| $d_8$ | 2.72048 | 0.03012 | 0.07075 |
| | Triclinic 20.6 GPa | | |
| $d_1$ | 2.03722 | 0.10725 | 0.32829 |
| $d_2$ | 2.15371 | 0.08585 | 0.20428 |
| $d_3$ | 2.14647 | 0.08257 | 0.25353 |
| $d_4$ | 2.14654 | 0.08254 | 0.25425 |
| $d_5$ | 2.42827 | 0.04946 | 0.11156 |
| $d_6$ | 2.43038 | 0.04926 | 0.11073 |
| $d_7$ | 2.53880 | 0.04048 | 0.09595 |
| $d_8$ | 2.64415 | 0.03485 | 0.09446 |